\documentclass[a4paper,12pt]{article}

\usepackage{jheppub}
\usepackage{cancel}
\usepackage{ragged2e}

\let\OLDthebibliography\thebibliography
\renewcommand\thebibliography[1]{
\OLDthebibliography{#1}
\setlength{\parskip}{0pt}
\setlength{\itemsep}{3.5pt plus 0.3ex}
}

\usepackage[english]{babel}
\usepackage{amsthm}
\newtheorem{remark}{Theorem}

\usepackage[T1]{fontenc}
\usepackage{amsmath}
\usepackage[percent]{overpic}
\usepackage{wrapfig}
\usepackage{tabu}
\usepackage{diagbox}

\usepackage{graphicx}
\usepackage{tikz}
\usepackage{import}
\usepackage{accents}
\usepackage{mathrsfs,amsmath,amssymb,slashed}
\usepackage{multirow,multicol}
\usepackage[percent]{overpic}
\usepackage{slashed}

\usepackage{wrapfig}
\usepackage{tabu}
\usepackage{diagbox}
\usepackage{mathrsfs,amsmath,amssymb,amsthm,amsfonts,tikz,graphicx,accents,hyperref}
\usepackage{leftidx}
\usepackage{import}
\usepackage{dsfont}

\DeclareFontFamily{OT1}{rsfs}{}
\DeclareFontShape{OT1}{rsfs}{m}{n}{ <-7> rsfs5 <7-10> rsfs7 <10->rsfs10}{} 

\DeclareMathAlphabet{\mycal}{OT1}{rsfs}{m}{n}
\def\mathbi#1{\textbf{\em #1}}

\newcommand{\bz}{\mathbf{z}}
\newcommand{\m}{\mathbi{m}}

\newcommand{\di}{\text{d}}

\newcommand{\cD}{\mathcal{D}}
\newcommand{\phii}{\varphi}

\newcommand{\Ad}{\text{Ad}}
\newcommand{\ad}{\text{ad}}
\newcommand{\der}{\partial}
\newcommand{\eps}{\varepsilon}

\newcommand{\be}{\begin{equation}}
\newcommand{\ee}{\end{equation}}
\newcommand{\Diff}{\text{Diff}\,S^1}
\newcommand{\Vect}{\text{Vect}\,S^1}
\newcommand{\ie}{\textit{i.e.}\ }
\newcommand{\eg}{\textit{e.g.}\ }

\newcommand{\bc}{\mathbf{c}}
\newcommand{\bC}{\mathbf{C}}
\newcommand{\CC}{\mathbb{C}}

\newcommand{\RR}{\mathbb{R}}
\newcommand{\ZZ}{\mathbb{Z}}

\DeclareMathOperator{\extdm}{d}
\newcommand{\extd}{\extdm \!}

\newcommand{\cA}{\mathcal{A}}

\newcommand{\cL}{\mathcal{L}}
\newcommand{\cM}{\mathcal{M}}
\newcommand{\cN}{\mathcal{N}}
\newcommand{\cO}{\mathcal{O}}
\newcommand{\cP}{P}
\newcommand{\cpz}{c}
\newcommand{\cQ}{Q}
\newcommand{\cT}{T}

\newcommand{\sfA}{\mathsf{A}}
\newcommand{\sfB}{\mathsf{B}}
\newcommand{\sfC}{\mathsf{C}}
\newcommand{\sfK}{\mathsf{K}}

\newcommand{\sfS}{\mathsf{S}}

\title{\bf{\textsf{Flat JT Gravity and the BMS-Schwarzian}}}

\author[]{Hamid Afshar$^{\,a}$ and Blagoje Oblak$^{\,b}$}

\affiliation[a]{\textit{Department of Physics, Faculty of Science, Ferdowsi University of Mashhad, Mashhad, Iran}}
\affiliation[b]{\textit{CPHT, CNRS, Ecole Polytechnique, IP Paris, F-91128 Palaiseau and LPTHE, Sorbonne Université, CNRS UMR 7589, F-75005 Paris (France)}}

\emailAdd{ham.afshar@gmail.com, blagoje.oblak@polytechnique.edu}

\abstract{We consider Minkowskian Jackiw-Teitelboim (JT) gravity in Bondi gauge at finite temperature, with non-zero vacuum energy. Its asymptotic symmetries span an extension of the warped Virasoro group, dubbed `BMS$_2$', which we investigate in detail. In particular, we show that this extension has a \textit{single} coadjoint orbit when central charges are real and non-zero. The ensuing BMS-Schwarzian action has no saddle points, and only coincides with the boundary action functional of flat JT gravity up to a crucial dilatonic zero-mode that ensures the existence of a well-defined bulk variational principle. We evaluate the corresponding gravitational partition function, which turns out to be one-loop exact precisely thanks to the presence of such a zero-mode.}

\begin{document}
\maketitle

\section{Motivation and outline}

The holographic relation between (quantum) gravity and lower-dimensional boundary theories has a long history by now \cite{Maldacena,WittenAdS}. In recent years, a surprisingly rich instance of this correspondence has come to be appreciated in two-dimensional (2D) dilaton-gravity theories \cite{Kitaev:15ur,Kitaev:2017awl,Almheiri:2014cka,Maldacena:2016hyu,Maldacena:2016upp,Engelsoy:2016xyb,Cvetic:2016eiv} such as Jackiw-Teitelboim (JT) gravity \cite{Jackiw:1984je,Teitelboim:1983ux} or the (gauged) Callan-Giddings-Harvey-Strominger model \cite{Callan:1992rs,Cangemi:1992bj}. These systems are indeed related to the low-energy effective theory describing the Sachdev-Ye-Kitaev (SYK) model \cite{Sachdev:1992fk,Sachdev:2010um,Kitaev:15ur,Kitaev:2017awl} and its generalizations. The core of the dictionary relies on symmetries: both 2D gravity and low-energy SYK admit a broken Virasoro symmetry under time reparametrizations, and their effective 1D action is a `Schwarzian'---the zero-mode of a conformal stress tensor on a coadjoint orbit of the Virasoro group \cite{Kirillov,Witten:1987ty,Stanford:2017thb}. This rephrasing has led to a flurry of activity aiming to understand the relation between SYK, quantum chaos and gravitational holography (see \eg \cite{Sarosi:2017ykf} and references therein for a pedagogical introduction).

Following this trend, various generalizations of SYK and its gravity dual have been proposed in the literature. A prominent recent example is the complex SYK model \cite{Sachdev:2015efa,Davison:2016ngz,Bulycheva:2017uqj,Chaturvedi:2018uov,Gaikwad:2018dfc,Gu:2019jub,Afshar:2019tvp}, whose low-energy limit yields a `refined' holographic correspondence sensitive to two chemical potentials rather than one. From the perspective of the two-dimensional (2D) bulk, this enhancement occurs thanks to a choice of fall-off conditions in Bondi gauge \cite{Afshar:2019axx,Godet:2020xpk,Afshar:2020dth,Godet:2021cdl} (see also the more recent papers \cite{Kar:2022sdc,Rosso:2022tsv}). The resulting asymptotic symmetries span an infinite-dimensional `Bondi-Metzner-Sachs' group in 2D \cite{Bondi,Sachs2}, or BMS$_2$ for short \cite{Afshar:2019axx}. However, all references dealing with this structure so far have eventually reduced it to its warped Virasoro subgroup \cite{Detournay:2012pc,Afshar:2015wjm}, either owing to dynamical equations of motion \cite{Afshar:2019axx,Afshar:2020dth}, or for convenience \cite{Godet:2020xpk}. The resulting boundary dynamics is then described by a warped Schwarzian action whose partition function is one-loop exact \cite{Afshar:2019tvp}, similarly to the standard Schwarzian theory of the Virasoro group \cite{Stanford:2017thb}.

The goal of this paper is to avoid this reduction to a subgroup and work with BMS$_2$ throughout. Indeed, it is the BMS$_2$ group, not warped Virasoro, that describes flat JT gravity in the sense that boundary gravitons transform under its coadjoint representation. It is therefore natural to take BMS$_2$ seriously and investigate the ensuing physics. The present work explores this question.

As we show, BMS$_2$ symmetry entails subtleties that do not normally occur in the boundary description of JT gravity, most notably the fact that the BMS-Schwarzian action functional has no saddle points. A hint of this behaviour already appears in the BMS$_2$ algebra, whose commutation relations in a basis of Fourier modes $(L_m,Q_m)$, respectively generating diffeomorphisms of Euclidean time and commuting `translations', read
\begin{align}
\label{t1b}
[L_m,L_n]
&=
(m-n)L_{m+n}+\frac{a}{12}m^3\delta_{m+n,0}\,,\\
\label{s1b}
[L_m,Q_n]
&=
-(m+n)Q_{m+n}+(bm-i\cpz)\delta_{m+n,0}\,,\\
\label{b1b}
[Q_m,Q_n]
&=
0\,,
\end{align}
where $a$ is a usual Virasoro central charge while $b$ and $\cpz$ are new. (These commutators were studied in \cite{Daniel-Jacob,Enriquez-Rojo:2021hna}; here we derive them in section \ref{sebalg}.) Note that $Q_0$ never appears on the right-hand side of \eqref{s1b}, since the $Q_m$'s are modes of a `current' with vanishing conformal weight. Furthermore, the central charge $\cpz$ is actually a zero-mode of gravitational boundary data and is not multiplied by any factor $m$ in \eqref{s1b}, so that $L_0$ and $Q_0$ fail to commute and span a Heisenberg subalgebra. This ultimately entails shift transformations of the translation current that turn the BMS$_2$ group itself into a single, giant coadjoint orbit, in contrast to orbits of the more standard Virasoro or warped Virasoro groups whose codimension is strictly positive. Despite this peculiarity of BMS$_2$, the gravitational bulk variational principle turns out to be well-defined, as the 1D version of the flat JT action functional differs from the BMS-Schwarzian by the addition of a crucial zero-mode. The latter is a dynamical quantity (it is an integration variable in the path integral), but fixing its value reduces the boundary action to the usual warped Schwarzian \cite{Afshar:2019axx,Afshar:2019tvp}, in accordance with a deeper relation between group structures. As a corollary, the partition function of the 1D theory is one-loop exact (even though it is \textit{not} a U(1) generator on a coadjoint orbit) and coincides with an integral of warped Schwarzian partition functions. We stress that, by contrast, there is no such thing as a `one-loop partition function' for the BMS-Schwarzian alone, as the latter has no saddle points.

The plan is as follows. First, section \ref{segrab} is purely gravitational. It is devoted to a brief review of JT gravity at finite temperature in Bondi gauge and the ensuing asymptotic symmetries, ending with a detailed derivation of its boundary action (eq.\ \eqref{Sell} below). We then introduce the BMS$_2$ group in section \ref{segg} and show that it generally contains three central charges, one of which happens to be the zero-mode of a gravitational boundary degree of freedom. We also show there that coadjoint orbits of BMS$_2$ are qualitatively different from those of other extensions of Diff$\,S^1$ \cite{Barnich:2015uva,Afshar:2015wjm}, as their codimension \textit{vanishes} (\ie their stabilizer is trivial up to central elements). The ensuing `BMS-Schwarzian' action functionals (eq.\ \eqref{s13t} below) have no saddle points and coincide with flat JT boundary actions up to a zero-mode. Finally, section \ref{sePAF} is devoted to the one-loop exact partition function of the boundary theory of flat JT gravity, computed in part thanks to the relation between BMS$_2$ and warped Virasoro. The \hyperref[app1]{appendices} contain further technical details on the adjoint representation of the BMS$_2$ group and its Maurer-Cartan form, respectively needed in sections \ref{segg} and \ref{sePAF}.

\section{JT gravity in Bondi gauge}
\label{segrab}

This section introduces our setup: Jackiw-Teitelboim gravity at finite temperature (including vacuum energy, with or without cosmological constant) with fall-offs imposed thanks to a Bondi gauge choice. As a result, asymptotic symmetries span a BMS$_2$ algebra \cite{Afshar:2019axx} and act on phase space in a linear way that will later (section \ref{segg}) be identified as a coadjoint representation. We end by deriving the 1D boundary action of the theory (eq.\ \eqref{Sell} below), which will be crucial for all subsequent sections.

\subsection{Asymptotic symmetries of JT gravity}
\label{segrav}

Consider a 2D manifold $\cM$ endowed with a Lorentzian metric $g$ with scalar curvature $R$, a real dimensionless scalar dilaton field $X$, and a vacuum energy density $\Lambda$.\footnote{$\Lambda$ does not affect space-time curvature, so we will \textit{not} refer to it as a cosmological constant (although this name is sometimes used in the literature) to stress its distinction from the actual cosmological constant, $-1/\ell^2$.} The associated \textit{Jackiw-Teitelboim (JT)} action functional with cosmological constant $-1/\ell^2$ reads
\be
\label{31}
I[g,X]
=
\frac{\kappa}{2}\int\extd^2 x\sqrt{|g|}\Big(X\big(R+2/\ell^2\big)-2\Lambda\Big)
\ee
where $\kappa$ is a dimensionless normalization (the inverse of Newton's constant in 2D). Our goal is to study the asymptotic symmetries of this theory in Bondi gauge, focussing especially on the Minkowskian limit where $\ell=\infty$. (Note that this is the only regime where $\Lambda$ matters: at finite $\ell$, vacuum energy can be cancelled by a constant shift of the dilaton.) The ensuing transformation law of boundary data is treated in section \ref{secops} at zero temperature, and in section \ref{sephatemp} at finite temperature. The additive boundary term needed to make the full action functional differentiable is built in section \ref{sebact}.

\paragraph{Symmetries preserve area.} Label the points of $\cM$ by Bondi coordinates $(u,r)$, respectively seen as retarded time and a `radius', and write the space-time metric in Bondi gauge:
\be
\di s^2
=
V(u,r)\di u^2-2\,\di u\,\di r\,,
\label{bondi}
\ee
where $V(u,r)$ can be any smooth function that behaves as $V(u,r)\sim-\tfrac{r^2}{\ell^2}+\cO(r)$ as $r\to+\infty$ at finite $u$. (We assume as usual that radial derivatives reduce the radial order of any function by at least one unit. In the flat limit, one just has $V=\cO(r)$.) We wish to find asymptotic Killing vector fields, that is, vector fields in space-time that preserve the gauge \eqref{bondi}. Any such vector field $\xi$ produces a Lie derivative such that $\cL_{\xi}g_{ur}=\cL_{\xi}g_{rr}=0$, giving the general expression
\be
\label{22}
\xi_{\eps,\eta}
=
\eps(u)\partial_u
-
\big(r\varepsilon'(u)+\eta(u)\big)\partial_r
\ee
in terms of unconstrained real functions $(\eps,\eta)$ on spatial or null infinity, with prime meaning $\der_u$. (The fall-offs of $V$ entail no further conditions on $(\eps,\eta)$.) Such vector fields span the \textit{BMS$_2$ algebra}, with Lie bracket 
\be
\label{liba}
[\xi_{\eps_1,\eta_1},
\xi_{\eps_2,\eta_2}]
=
\xi_{[\eps_1,\eps_2],
(\eps_1\eta_2)'-(\eps_2\eta_1)'}
\ee
where $[\eps_1,\eps_2]\equiv\eps_1\eps_2'-\eps_2\eps_1'$ is the usual bracket of 1D vector fields. The functions $(\eps,\eta)$ in \eqref{22} thus have definite weights under reparametrizations of time: $\eps(u)\der_u$ is a vector field, while $\eta(u)\di u$ is a one-form.\footnote{We abusively write $\eps$ for both the vector field $\eps(u)\der_u$ and its component $\eps(u)$; the same applies to $\eta$.} 
Geometrically, $\eps$ generates diffeomorphisms of time while $\eta$ produces time-dependent radial `translations'. This can be made manifest by exponentiating any vector field of the form \eqref{22} into a finite diffeomorphism
\be
\label{s3t}
(u,r)
\longmapsto
\Big(f(u),\frac{r}{f'(u)}-\eta\big(f(u)\big)\Big)
\ee
where $f$ is an orientation-preserving diffeomorphism of the real line (so $f'(u)>0$ for all $u$). Note that any such transformation preserves the area form\footnote{Incidentally, area-preserving diffeomorphisms of the form \eqref{s3t} with $\eta=0$ (and Euclidean $(u,r)$) appear in the quantum Hall effect, where they span the Virasoro symmetry of gapless edge modes \cite{Cappelli:1994wb}.} $\di u\wedge \di r$, as must indeed be the case owing to the gauge choice \eqref{bondi}. This holds regardless of the cosmological constant $-1/\ell^2$.

\paragraph{Warped Virasoro.} It is worth noting that the algebra \eqref{22}--\eqref{liba} is very close to (but emphatically different from) the usual symmetry of warped CFTs \cite{Detournay:2012pc,Afshar:2015wjm}. Indeed, consider the subgroup of BMS$_2$ whose one-forms are exact: $\eta=\di\sigma$ for some function $\sigma$. The bracket \eqref{liba} then becomes $\big[(\eps_1,\sigma_1'),(\eps_2,\sigma_2')\big]=\big([\eps_1,\eps_2],(\eps_1\sigma'_2-\eps_2\sigma'_1)'\big)$, which reduces to the usual commutator of warped conformal transformations \cite{Detournay:2012pc} up to a total derivative that makes the zero-mode of $\sigma$'s irrelevant. We shall return to this algebraic relation in section \ref{sephatemp}, where generators will be defined in terms of Fourier modes on a thermal cycle. The embedding of warped Virasoro in BMS$_2$ will leave traces throughout this work, including the crucial reduction of the BMS-Schwarzian (section \ref{segg}) to the warped Schwarzian action \cite{Afshar:2019axx,Afshar:2019tvp,Godet:2020xpk}.

\subsection{Covariant phase space and its transformation laws}
\label{secops}

Our considerations were purely kinematical so far, but it is essential for future reference to keep in mind that BMS$_2$ acts on (covariant) phase space, \ie on the set of on-shell field configurations specified by the JT action \eqref{31}. We now explore this space and its transformations under BMS$_2$.

\paragraph{Phase space.} The equations of motion due to the JT action \eqref{31} read (we use $\approx$ for on-shell equalities)\footnote{The variation of \eqref{31} contains a boundary term that does not affect equations of motion, so we bluntly discard it for now. We return to a proper construction of the differentiable improved action in section \ref{sebact}.}
\be
\label{eom}
R
\approx
-2/\ell^2,
\qquad
\nabla_\mu\nabla_\nu X -g_{\mu\nu}\nabla^2 X+\frac{1}{\ell^2}g_{\mu\nu}X
\approx
g_{\mu\nu}\Lambda\,.
\ee
In Bondi gauge \eqref{bondi}, one has $R=\der_r^2V$ and the $rr$ component of \eqref{eom} sets $\der_r^2X\approx0$, so
\be
\di s^2
\approx
-\Big(\frac{r^2}{\ell^2}+2\cP(u)r+2\cT(u)\Big)\di u^2-2\,\di u\,\di r\,,
\qquad
X\approx x(u)r+y(u)\,,
\label{metric}
\ee
where $\cT$, $\cP$, $x$ and $y$ are arbitrary real functions of time at this stage. The remaining components of \eqref{eom} then provide genuine dynamics for $(x,y)$, namely the linear evolution equations
\be
\label{dotx}
x'-\cP x-\Lambda+\frac{y}{\ell^2}
\approx
0\,,
\qquad
y''+\cP y'-\cT'x-2\cT x'
\approx
0\,.
\ee
Points in covariant phase space are thus, apparently, labelled by two arbitrary functions $(\cT,\cP)$, along with initial conditions $(x(0),y(0),y'(0))$ for eqs.\ \eqref{dotx}. Note the odd number of initial conditions, contradicting the fact that any symplectic manifold is even-dimensional; this will eventually be cured in section \ref{sebact}, where the existence of a well-defined variational principle will impose an additional constraint on $x,y$, effectively removing one initial condition and making the dilatonic phase space two-dimensional. 

\paragraph{Transformation laws.} Since any asymptotic Killing vector field \eqref{22} preserves the Bondi gauge \eqref{bondi}, it also maps any on-shell metric \eqref{metric} on another metric of the same form, yielding a well-defined transformation law for `boundary gravitons' $(\cT,\cP)$. Evaluating the relevant Lie derivatives of the metric, one finds
\be
\label{28}
\delta\cT
=
\eps\cT'+2\eps'\cT-\eta\cP-\eta'\,,
\qquad
\delta\cP
=
\eps\cP'
+\eps'\cP
-\eps''
-\frac{\eta}{\ell^2}\,.
\ee
When $\eta=\sigma'$ is a total derivative, this coincides with the coadjoint representation of the warped Virasoro algebra with a stress tensor $\cT$ and a `momentum density' $\cP$ \cite{Afshar:2015wjm}. For arbitrary $\eta$, however, one expects \eqref{28} to reproduce the coadjoint representation of BMS$_2$, which will indeed turn out to be the case in the flat limit $\ell\to\infty$ (see sections \ref{sephatemp} and \ref{serta}).\footnote{Since $\ell$ is dimensionful, the limit is actually taken on dimensionless quantities---see \cite{Barnich:2012aw} for details. We simply write this as $\ell\to\infty$ for brevity.} The asymptotic Killing vector fields \eqref{22} also preserve the form \eqref{metric} of the dilaton: computing $\delta X=\xi^{\mu}\der_{\mu}X$, one finds the transformation law of on-shell functions ($x,y$),
\be
\label{s4b}
\delta x
=
\eps x'-\eps' x\,,
\qquad
\delta y
=
\eps y'
-\eta x\,.
\ee
This suggests that $x$ is a vector field while $y$ is a function, up to an unfamiliar shift transformation by $\eta$. In fact, we will see in section \ref{sebogg} that eqs.\ \eqref{s4b} reflect the action of a left-invariant vector field living on the BMS$_2$ group manifold (and specified by $(\eps,\eta)$) on a BMS$_2$ group element (specified by $(x,y)$). From that perspective, the pair $(\cT,\cP)$ that determines a gravitational background \eqref{metric} is essentially an element of the dual of the BMS$_2$ algebra, while $(x,y)$ labels a point in (a quotient of) the BMS$_2$ group.

Note that asymptotic Killing vectors trivially include \textit{exact} Killing vector fields, \ie generators of isometries. In that context, there exists a sharp difference between AdS backgrounds and flat backgrounds. Requiring $\delta\cP=0$ in \eqref{28} yields indeed $\eta=\ell^2[(\eps\cP)'-\eps'']$, whereupon the transformation law of $\cT$ reduces to that of the CFT stress tensor $\cT-\ell^2(\cP'+\cP^2/2)$ under usual conformal transformations. Isometries at finite $\ell$ thus coincide with typical CFT stabilizers; at finite temperature, those can only be one-dimensional Abelian groups or (covers of) SL$(2,\RR)$ \cite{Witten:1987ty}. No such simplification occurs in the flat limit $\ell=\infty$, where the stabilizer condition $\delta\cT=\delta\cP=0$ is a pair of non-trivial differential equations for $(\eps,\eta)$. This is consistent with the expectation that isometries of flat backgrounds should \textit{not} span the same groups as those of AdS backgrounds. We return to isometries at the end of the next subsection.

\subsection{Phase space at finite temperature}
\label{sephatemp}

Most of this work will be concerned with \textit{Euclidean} JT gravity at finite temperature $1/\beta$. Euclidean time $\tau=iu$ then becomes $\beta$-periodic and the space-time boundary at infinity ($r\to\infty$) is a circle whose points can be labelled by an angle $\phii=2\pi\tau/\beta\sim\phii+2\pi$. The Bondi gauge metric \eqref{bondi} becomes $\di s^2=V(\phii,r)\di\phii^2-2(\beta/(2\pi i))\di\phii\,\di r$ for some function $V\sim\tfrac{\beta^2r^2}{(2\pi\ell)^2}+\cO(r)$, so that the (singular) flat limit coincides with the regime of high temperatures in AdS$_2$. We now describe the corresponding symmetries and phase space.

\paragraph{Asymptotic symmetries.} Asymptotic Killing vector fields at finite temperature still take the form \eqref{22}, now with functions $(\eps,\eta)$ that are $2\pi$-periodic in $\phii$. In fact, owing to the respective conformal weights $(-1,1)$ of $(\eps,\eta)$, we adopt the convention that the Wick rotation of \eqref{22} is
\be
\xi_{\eps,\eta}
\equiv
\eps(\phii)\partial_{\phii}
-\Big(r\eps'(\phii)+\frac{2\pi i}{\beta}\eta(\phii)\Big)\partial_r\,,
\label{tt3b}
\ee
where $\eps$ is real and $\eps'\equiv\partial_{\phii}\eps$. The bracket of such vector fields is again given by eq.\ \eqref{liba}, now with $'\equiv\partial_{\phii}$. Accordingly, a Fourier basis of the (complexified) BMS$_2$ algebra is
\be
\label{s3q}
\ell_m\equiv-i\,\xi_{e^{im\phii},0}\,,
\qquad
q_m\equiv-i\,\xi_{0,e^{im\phii}}
\qquad(m\in\ZZ),
\ee
whereupon the Lie bracket \eqref{liba} yields the centreless form of eqs.\ \eqref{t1b}--\eqref{b1b},
\be
\label{libam}
[\ell_m,\ell_n]
=
(m-n)\ell_{m+n}\,,
\qquad
[\ell_m,q_n]
=
-(m+n)q_{m+n}\,,
\qquad
[q_m,q_n]
=
0\,.
\ee
The centreless warped Virasoro algebra alluded to in section \ref{segrav} is then obtained from \eqref{libam} by discarding $q_0$ and defining $p_m\equiv mq_m$ for all $m$, including the convention $p_0\equiv0$ for the Casimir of warped translations. Also note the 2D Euclidean (complexified Poincar\'e) subalgebras of \eqref{libam} generated by the rotation (or boost) $\ell_0$ and the `translations' $\ell_n,q_{-n}$ with some fixed $n\neq0$. These will turn out to be the isometries of $n$-fold covers of the Euclidean plane.

\paragraph{Covariant phase space.} Similarly to asymptotic Killing vectors \eqref{tt3b}, one can Wick-rotate the on-shell functions $(\cT,\cP,x,y)$ according to their conformal weights so that the Euclidean version of eqs.\ \eqref{metric} reads
\be
\label{214}
\di s^2
\approx
\Big(\frac{\beta^2r^2}{(2\pi\ell)^2}
-2r\frac{\beta}{2\pi i}\cP(\phii)
-2\cT(\phii)
\Big)\di\phii^2
-2\frac{\beta}{2\pi i}\di\phii\,\di r\,,
\qquad
X
\approx
\frac{\beta}{2\pi i}x(\phii)r+y(\phii)\,.
\ee
Here all four functions $(\cT,\cP,x,y)$ are $2\pi$-periodic, and generally complex: for instance, the metric of the $n$-fold cover of the Euclidean plane is obtained with $\cP=-in$ and $\cT=0$. Reality conditions will eventually be chosen in section \ref{secomp} to set up suitable path integration contours.

Up to the replacement of $u$ by $\phii$ and $'\equiv\partial_{\phii}$, the transformation law of the Wick-rotated variables $(\cT,\cP,x,y)$ under asymptotic symmetries \eqref{tt3b} is still given by eqs.\ \eqref{28}--\eqref{s4b}. Furthermore, the dilaton's equations of motion still take the form \eqref{dotx} with the same replacements. (This coincidence between Lorentzian and Euclidean formulas is due to the fact that Wick rotations were chosen in accordance with conformal weights, similarly to the usual formulas for the Wick rotation of spinful fields.)

As mentioned in section \ref{secops}, one expects the transformation law \eqref{28} of boundary gravitons to be related to the coadjoint representation of the BMS$_2$ algebra. To anticipate this, note from \eqref{28} that $\cT$ transforms under reparametrizations as a primary with weight two, while $\cP$ is a one-form. Splitting $\cP(\phii)=\cpz+\cQ'(\phii)$ into a zero-mode $\cpz$ and a total derivative, the flat version ($\ell=\infty$) of \eqref{28} leaves $\cpz$ invariant and suggests the transformation law
\be
\label{ifit1}
\delta\cT
=
\eps\cT'+2\eps'\cT
-\eta (\cQ'+\cpz)
-\eta'\,,
\qquad
\delta\cQ
=
\eps(\cQ'+\cpz)
-\eps'
\ee
for the pair $(\cT,\cQ)$. We show in section \ref{serta} that this is the coadjoint representation of the BMS$_2$ algebra with two central charges, one of which is the constant $\cpz$. (One could of course add any constant to the right-hand side of $\delta\cQ$ in \eqref{ifit1} without affecting $\delta\cP$, but it turns out that it is precisely \eqref{ifit1} that coincides with the coadjoint representation of BMS$_2$.) Note that these conventions relate the zero-modes $\cT_0$ and $\cpz$ to the temperature and Rindler horizon radius $r_{\text{H}}$ according to $2\pi i\,\cT_0/\cpz=-\beta\,r_{\text{H}}$. This combination turns out to be important: we shall see in section \ref{secomp} that $2\pi\cT_0/\cpz$ is essentially the on-shell Euclidean action that yields the tree-level term of the partition function.\footnote{In fact, the on-shell action of Euclidean Flat JT gravity is $S^{(0)}=\kappa\Lambda\beta (r_0+r_{\text{H}})$, where $r_0=\int_0^{r_0}\extd r$ is an infrared-divergent constant. We will not discuss possible counterterms making the on-shell action finite.}

\paragraph{Isometries of Euclidean flat metrics.} To develop some intuition on the space of metrics \eqref{214}, one may investigate their isometry algebras. As mentioned in section \ref{secops}, these are just stabilizer algebras of Virasoro coadjoint orbits at finite AdS radius $\ell$. At infinite $\ell$, the situation is more involved. Accordingly, let $(\cT,\cP)=(\cT_0,\cpz)$ be constants and set $\delta\cT=\delta\cP=0$ in eqs.\ \eqref{28}, assuming $\cpz\neq0$ for now. This yields differential equations for $(\eps,\eta)$ whose general solution is
\be
\eps(\phii)
=
A\,e^{c\,\phii}+B+\cO(1/\ell^2)\,,
\qquad
\eta(\phii)
=
A\cT_0\,e^{c\,\phii}+Ce^{-c\,\phii}+\cO(1/\ell^2)\,.
\label{s66b}
\ee
Requiring $2\pi$-periodicity sets almost all integration constants to zero when $\cpz$ is real, and one is left with $\eps=\text{cst}$, $\eta=0$. There is much more leeway when symmetries and boundary gravitons are complex: then the functions $(\eps,\eta)$ in \eqref{s66b} are $2\pi$-periodic whenever $c\in i\,\ZZ^*$, in which case the space of BMS$_2$ vector fields given by \eqref{s66b} coincides with a complexified Euclidean algebra of the form mentioned below \eqref{libam}: the corresponding bulk vector fields \eqref{tt3b} are
\be
\xi_0
=
\partial_\varphi\,,
\quad
\xi_1
=
e^{c\,\varphi}\partial_\varphi-e^{c\,\varphi}\Big(r\,c +\frac{2\pi i}{\beta}T_0\Big)\partial_r\,,
\quad
\xi_2
=
-\frac{2\pi i}{\beta}e^{-c\,\varphi}\partial_r
\ee
and may respectively be identified with $(-i\ell_0,\ell_n,q_{-n})$ in terms of the modes \eqref{s3q} when $c=in$. The gravitational background having these isometries is an $n$-fold cover of the Euclidean plane (\ie a thermal disk, generally with a quantized conical excess at the origin). In the exceptional case $\cpz=0$, $2\pi$-periodic isometry generators are given by $\eps=\text{cst}$, $\eta=\text{cst}'$ instead of eqs.\ \eqref{s66b}, generating the usual isometry group $\RR\times\text{U}(1)$ of a cylinder; however, to the extent that $\cpz\neq0$ will be a crucial condition in what follows, the cylinder eventually turns out to be excluded from the phase space considered in this work.

\subsection{Boundary action of flat JT gravity}
\label{sebact}

As in any field theory, defining the phase space of JT gravity requires a choice of fall-off conditions. These, in turn, typically imply that the bulk action functional \eqref{31} is not differentiable, \ie that the equations of motion \eqref{eom}--\eqref{dotx} do not provide its true extremum. We glossed over this key detail in section \ref{secops}, so we now rectify the omission---in great part because the resulting boundary term recasts the entire theory as a Schwarzian-like 1D system whose (Euclidean) action will be studied in depth in the remainder of this work. In particular, the boundary action turns out to be closely related to BMS-Schwarzians to be introduced in section \ref{segg}. Note that the argument that follows closely parallels the derivation of \cite[sec.\ 2.1]{Godet:2020xpk}, except that we include the effects of the vacuum energy $\Lambda$ (which are non-trivial in the flat limit $\ell\to\infty$).

\paragraph{Improved action.} As before, consider the space-time manifold $\cM$ (Lorentzian for now) endowed with a metric $g$ and a dilaton $X$ governed by the bulk action \eqref{31}. The latter must generally be supplemented by a boundary term so as to ensure the existence of a well-defined variational principle. Indeed, upon fixing Bondi gauge \eqref{bondi} and assuming (off-shell) large $r$ fall-offs of the form $V(u,r)\sim-\tfrac{r^2}{\ell^2}-2\cP(u)r-2\cT(u)+o(1)$, along with the dilaton fall-offs $X(u,r)\sim x(u)r+y(u)+o(1)$, the variation of the bulk action \eqref{31} reads
\be
\label{vari}
\delta I
=
\kappa\int\di^2x\,(\text{EOM})
+\kappa\int\di u\,\big(x\,\delta\cT-y\,\delta\cP\big)\,,
\ee
where `EOM' denotes a term that vanishes when the equations of motion \eqref{eom} are satisfied. The boundary term in \eqref{vari} makes it manifest that the functional \eqref{31}, by itself, is \textit{not} differentiable, so that the equations of motion \eqref{eom} do \textit{not} provide its true extremum.

In order to obtain a differentiable action (or at least one that is differentiable on-shell), one must add a boundary term to the bulk functional \eqref{31}. Since the fields $(\cT,\cP)$ and $(x,y)$ are independent off-shell, \eqref{vari} shows that they must, somehow, be related in order for a suitable boundary action to cancel the term $\int(x\delta\cT-y\delta\cP)$. A common solution to this problem in 2D dilaton gravity is to use the equations of motion \eqref{dotx}. We stress that this is somewhat exotic compared to other boundary actions in physics---most notably Chern-Simons theory \cite{Elitzur,Coussaert}, where one requires the action functional to be differentiable \textit{everywhere} in a space of fields subject to fall-offs. Here, by contrast, we shall require the action to be merely differentiable \textit{at on-shell configurations}. This ensures that bulk equations of motion provide true extrema of the action, which is enough for the path integral to admit a well-defined semi-classical regime.

Accordingly, let the full action of the system be $S_{\text{full}}[g,X]=I[g,X]+S[g,X]$, where $I$ is the bulk action \eqref{31}. Our goal is to find a boundary term $S$ such that $\delta S_{\text{full}}\approx0$ without any boundary contribution. Owing to the variation \eqref{vari}, one must have
\be
\label{deltab}
\delta S
\approx
-\kappa\int\di u\,(x\,\delta\cT-y\,\delta\cP)\,,
\ee
so it seems natural to attempt $S\propto\int(\cT x-\cP y)$. This is not enough, however, as $\delta S$ then contains variations of $(x,y)$ that do not appear in \eqref{deltab}; the cure is readily found thanks to the equations of motion \eqref{dotx}, which imply that the combination
\be
\label{s5b}
xy'-\cT x^2-\Lambda y+\frac{y^2}{2\ell^2}
\ee
is \textit{constant} ($u$-independent) on-shell, so that the boundary term
\be
\label{BBB}
S
\equiv
-\kappa
\int\di u\,
\Big(\cT x-\cP y+\frac{x'}{x}y-\Lambda\frac{y}{x}+\frac{y^2}{2\ell^2 x}\Big)
\ee
has an on-shell variation
\be
\delta S
\approx
-\kappa\int\di u\,(x\delta\cT-y\delta\cP)
-\kappa\Big(xy'-\cT x^2-\Lambda y+\frac{y^2}{2\ell^2}\Big)
\,\delta\left[\int\frac{\di u}{x}\right]\,.
\label{delb}
\ee
(Here we have neglected a total time derivative, which is harmless as usual.) This nearly satisfies the requirement \eqref{deltab}: the only issue is the extra variation of the integral of $1/x$. In both Lorentzian and Euclidean signatures, this is settled by requiring that $\int\di u/x$ be finite and fixed once and for all. At finite temperature, one thus declares that $x(\phii)\neq0$ at all Euclidean times $\phii$ and that the integral
\be
\label{zemo}
\oint_0^{2\pi}
\frac{\di\phii}{x(\phii)}
\equiv
\cA
\ee
is some finite parameter that does \textit{not} vary when computing the equations of motion.  One may think of this as an extra, off-shell boundary condition on the dilaton \cite{Gonzalez:2018enk,Afshar:2019axx}.

Provided the zero-mode of $1/x$ is fixed, off-shell, according to eq.\ \eqref{zemo}, the improved action $S_{\text{full}}=I+S$ does reach its extremum at on-shell field configurations (thanks to eq.\ \eqref{delb} and the constancy of \eqref{s5b}). Since equations of motion were explicitly used to derive \eqref{delb}, the full action is differentiable \textit{on-shell}; no claim is made as to its off-shell differentiability. In any event, the action $S_{\text{full}}$ evaluated on metrics that satisfy the constraint $R\approx-2/\ell^2$ reduces to the pure boundary term \eqref{BBB} up to an irrelevant bulk volume contribution $\propto\Lambda$. Eq.\ \eqref{BBB} thus coincides, for all practical purposes, with the boundary action of JT gravity, and one verifies indeed that its equations of motion reproduce \eqref{dotx}.

\paragraph{The dilaton as a group element.} For later comparison with BMS-Schwarzian actions, it is useful to rewrite the Euclidean version of the boundary action \eqref{BBB} in terms of group elements. Accordingly, let $\phii\equiv2\pi\tau/\beta$ be rescaled Euclidean time and implement the constraint \eqref{zemo} by declaring
\be
\label{s6t}
\frac{1}{f'(\phii)}
\equiv
\frac{\cA}{2\pi}x(\phii)
\ee
where $f(\phii)$ is a real function that satisfies $f'(\phii)>0$ and $f(\phii+2\pi)=f(\phii)+2\pi$ for all $\phii$. (Thus $f$ is really a diffeomorphism of the circle: we return to this in section \ref{segg}.) As for $y$, write it as a function $\eta$ given by
\be
\label{s6tt}
\eta(\phii)
\equiv
-\frac{\cA}{2\pi}y(f^{-1}(\phii))
\ee
where $f^{-1}$ is the inverse of the diffeomorphism $f$. Note that the off-shell condition \eqref{zemo} is subject to a dynamical constraint: the first equation of motion in \eqref{dotx} implies
\be
\label{s6b}
\Lambda\cA
=
-2\pi\left(\cpz+\frac{\eta_0}{\ell^2}\right)
\ee
where $\cpz\equiv\tfrac{1}{2\pi}\oint\cP$ is the zero-mode of $\cP$ as in \eqref{ifit1} and $\eta_0\equiv\tfrac{1}{2\pi}\oint\eta$ is the zero-mode of $\eta$. At finite $\ell$, this suggests that $\eta_0$ should be fixed off-shell when $\cpz$ and $\cA$ are fixed, which is indeed the choice typically made in (Bondi gauge) JT gravity on AdS$_2$ and eventually leads to the warped Schwarzian theory \cite{Afshar:2019axx,Godet:2020xpk,Afshar:2020dth}. By contrast, in the flat limit $\ell\to\infty$, eq.\ \eqref{s6b} does not involve $\eta$, which is left unconstrained.

We have now gathered all the ingredients needed to express the boundary action of JT gravity in terms of group elements. The Euclidean form of the action \eqref{BBB} with the replacements \eqref{s6t}--\eqref{s6tt} thus becomes
\be
\label{Sell}
S
=
\frac{2\pi i\kappa}{\cA}
\left\{%
\oint\frac{\di\phii}{f'(\phii)}
\bigg[%
\cT+\Big(\big(\cpz+\cQ'\big)\,f'
+f''
-\cpz f'^2\Big)\eta\circ f\bigg]
+\oint\di\phii\frac{\eta^2-2\eta_0^2}{2\ell^2}\right\}\,,
\ee
where we chose to write the boundary graviton $\cP$ as $c+\cQ'$. At finite $\ell$, the zero-mode of $\eta$ is fixed according to \eqref{s6b}, so the term $\propto\cpz f'^2\eta\circ f$ is irrelevant. One can then write $\eta=\eta_0+\sigma'$ for some periodic function $\sigma$, whereupon \eqref{Sell} reduces to the warped Schwarzian action \cite{Afshar:2019axx}. This is consistent with the correspondence between warped Virasoro and BMS$_2$ sketched in sections \ref{segrav}--\ref{secops}. The situation is completely different in the flat limit $\ell=\infty$, where $\eta_0$ is an unconstrained Lagrange multiplier. The term $\propto\cpz f'^2\eta\circ f$ then becomes important: it is in fact responsible for the very existence of a saddle point.

The remainder of this paper is devoted to a detailed investigation of the action functional \eqref{Sell} in the strict flat limit $\ell=\infty$. In particular, we show in the next section that it is a Schwarzian action of BMS$_2$, up to the zero-mode of $\eta$. Note that the coefficient $\propto i\kappa/\cA$ in front of the action is actually real when $\cpz$ and $\eta_0$ are purely imaginary, owing to the relation \eqref{s6b}. Indeed, we saw around eq.\ \eqref{214} that $\cP=\cpz+\cQ'$ is purely imaginary in order to include the Euclidean plane in the gravitational phase space, while the condition $\eta\in i\RR$ will be enforced in section \ref{secomp} in order for the path integral of the thermal theory to converge.

\section{The \texorpdfstring{BMS$_2$}{BMS2} group and BMS-Schwarzian actions}
\label{segg}

We have seen that the BMS$_2$ algebra describes asymptotic symmetries of JT gravity in Bondi gauge (with or without cosmological constant), and that the Euclidean boundary action \eqref{Sell} can be written in Schwarzian-like form, similarly to other known examples in dilaton-gravity theories (see \eg the sample \cite{Kitaev:15ur,Kitaev:2017awl,Almheiri:2014cka,Maldacena:2016hyu,Maldacena:2016upp,Afshar:2019axx,Godet:2020xpk,Afshar:2020dth}). The goal of this section is therefore (i) to describe in detail the structure of the centrally extended BMS$_2$ group, along with its adjoint and coadjoint representations, and (ii) to deduce the corresponding Schwarzian-like action functionals. As we shall see, coadjoint orbits of BMS$_2$ are peculiar in that their codimension vanishes: there is a single orbit, diffeomorphic to the centreless BMS$_2$ group itself, for all non-zero real values of BMS$_2$ central charges. As a result, BMS-Schwarzian actions have no saddle points, contradicting the naive expectation that \eqref{Sell} (which does have saddle points) is in fact a Schwarzian. This tension is resolved by the fact that \eqref{Sell} with $\ell=\infty$ is actually a BMS-Schwarzian \textit{plus} the crucial zero-mode of $\eta$, which reinstates the presence of saddle points.

Note that this entire section is focussed on the Euclidean theory at finite temperature, so that group and algebra elements are functions of a periodic coordinate $2\pi\tau/\beta\equiv\phii\sim\phii+2\pi$. Furthermore, we mostly restrict attention to the \textit{real} form of the BMS$_2$ group. Aspects of complexification will only be addressed in passing for now, postponing a more careful treatment to the path integral considerations of section \ref{secomp}.

\subsection{The \texorpdfstring{BMS$_2$}{BMS2} group}
\label{sebogg}

Here we provide a detailed definition of the BMS$_2$ group and its central extensions. As it turns out, BMS$_2$ admits three central charges, two of which are realized in flat JT gravity (and implicitly appear \eg in the transformation laws \eqref{ifit1}). This is a key prerequisite for the upcoming considerations on the BMS$_2$ algebra (section \ref{sebalg}), on BMS$_2$ coadjoint orbits (section \ref{serta}), and on their Schwarzian actions (section \ref{seBeSch}).

\paragraph{Centreless group.} Consider the group $\Diff$ of (orientation-preserving) diffeomorphisms of the circle, whose elements are real functions $f$ of an angle $\phii\in\RR$ such that $f'(\phii)>0$ and $f(\phii+2\pi)=f(\phii)+2\pi$ for all $\phii$. (We anticipated this definition below \eqref{s6t}.) Consider also the vector space $\Omega^1(S^1)$ of one-forms on the circle, written as $\eta=\eta(\phii)\di\phii$ in terms of their $2\pi$-periodic component $\eta(\phii)$. Diffeomorphisms act on one-forms by pullback, defined as $(f^*\eta)(\phii)\di\phii=\eta(f(\phii))f'(\phii)\di\phii$. In CFT language, this says that $\eta$ is a primary field with unit weight (a current), in accordance with the Lie bracket \eqref{liba}. The \textit{centreless BMS$_2$ group} then is a semi-direct product $\Diff\ltimes\Omega^1(S^1)$, whose elements are pairs $(f,\eta)$, with multiplication
\be
(f_1,\eta_1)\cdot(f_2,\eta_2)
\equiv
\big(%
f_1\circ f_2
,\,
\eta_1+ (f_1^{-1})^*\eta_2
\big)\,.
\label{mu}
\ee
Following the terminology introduced above \eqref{s3t}, we refer to $f$ as a diffeomorphism or a reparametrization (of retarded time), while $\eta$ is a `translation'. As in higher-dimensional BMS groups, translations span an Abelian normal subgroup of BMS$_2$.

Despite its simplicity, the group structure \eqref{mu} already contains information that can be matched with gravitational observations in section \ref{segrab}. For instance, the centreless BMS$_2$ algebra consists of pairs $(\eps,\zeta)$, where $\eps=\eps(\phii)\partial_{\phii}$ is a vector field while $\zeta$ is again a one-form, as in the asymptotic Killing vector fields \eqref{22}--\eqref{tt3b}. (The bracket \eqref{liba} will similarly be recovered in section \ref{sebalg}.) As in any Lie group, the Lie algebra can be seen as the tangent space at the identity, or, equivalently, as the space of left-invariant vector fields on the group manifold \cite[sec.\ 4.1]{Abraham}. Each pair $(\eps,\zeta)$ thus defines a left-invariant vector field on BMS$_2$ acting on a point $(f,\eta)$ according to
\be
\label{ss8b}
(\delta f,\delta\eta)
\equiv
\frac{\partial}{\partial t}\bigg|_{t=0}\Big[(f,\eta)\cdot(e^{t\eps},t\zeta)\Big]
\stackrel{\text{\eqref{mu}}}{=}
\big(\eps f',\zeta\circ f^{-1}\,(f^{-1})'\big)\,,
\ee
from which it follows that the combinations $x\equiv1/f'$ and $y\equiv-\eta\circ f$ transform under BMS$_2$ exactly as the components of the dilaton in eqs.\ \eqref{s4b}, up to renaming $\eta$ as $\zeta$. This confirms that the earlier identifications \eqref{s6t}--\eqref{s6tt} are indeed compatible with the BMS$_2$ group structure.

As an aside, note that warped conformal transformations \cite{Detournay:2012pc} span a subgroup of \eqref{mu} obtained by declaring that all BMS$_2$ translations are total derivatives ($\eta=\sigma'$). Eq.\ \eqref{mu}, however, only provides a realization of this `warped $\Diff$ group' in which the Casimir generating constant translations is set to zero. Non-zero Casimirs turn out to be allowed only when central charges are switched on.

\paragraph{Central extensions.} We have already encountered two central charges, implicitly at least, in the inhomogeneous terms of eqs.\ \eqref{ifit1}. It is therefore essential to classify the possible central extensions of BMS$_2$. To do this we follow \cite{Ovsienko:1994im}, where extensions of the BMS$_2$ algebra were classified on a cohomological basis; here we describe the corresponding group extensions.

Consider therefore the set $\Diff\times\Omega^1(S^1)\times\RR^3$, whose elements $(f,\eta;\,\bz)$ extend our earlier notation by a three-component\footnote{The presence of exactly \textit{three} central terms, and no more, stems from Lie algebra cohomology \cite{Ovsienko:1994im}. The variables $x,y$ are just numbers and have nothing to do with the components of the dilaton introduced in \eqref{metric}. That they are real is merely a choice: one is free to consider complex central charges as well.} central element $\bz=(x,y,z)\in\RR^3$. We endow this set with the following binary operation, extending the multiplication \eqref{mu}:
\be
(f_1,\eta_1;\,\bz_1)\cdot(f_2,\eta_2;\,\bz_2)
\equiv\Big(
f_1\circ f_2,\eta_1+ (f_1^{-1})^*\eta_2;\,\bz_1+\bz_2+\bC[f_1,f_2,\eta_2]
\Big)\,,
\label{muc}
\ee
where $\bC[f_1,f_2,\eta_2]=\big(\sfA[f_1,f_2], \sfB[f_1,\eta_2], \sfC[f_1,\eta_2]\big)\in\RR^3$ is a triple of real cocycles given by
\begin{align}
\label{coc}
\sfA[f_1,f_2]
&\equiv
-\frac{1}{48\pi}\oint\di\phii\log\,\big(f_1'(f_2(\phii))\big)\,\frac{f_2''(\phii)}{f_2'(\phii)}\,,\\
\label{cok}
\sfB[f,\eta]
&\equiv
\frac{1}{2\pi}\oint\log\big(f'(\phii)\big)\,\eta(\phii)\,\di\phii\,,\\
\label{coq}
\sfC[f,\eta]
&\equiv
\frac{1}{2\pi}\oint
\big(f(\phii)-\phii\big)\,\eta(\phii)\,\di\phii\,,
\end{align}
all integrals being taken over $\phii\in[0,2\pi]$. The fact that these are \textit{cocycles} ensures that \eqref{muc} is associative, defining the \textit{centrally extended BMS$_2$ group}. In particular, the functional \eqref{coc} is the standard Virasoro-Bott cocycle \cite{Bott,Guieu}, while \eqref{cok}--\eqref{coq} appear to be new. Eq.\ \eqref{cok}, for instance, pairs the one-form $\eta$ with the one-cocycle $\log f'$; it is similar in this sense to the BMS$_3$ extension that pairs supertranslations with the Schwarzian derivative of superrotations \cite{Barnich:2014kra}. The last cocycle, eq.\ \eqref{coq}, is somewhat atypical in that it involves no derivatives; it will eventually turn out to be crucial for 2D gravity, as its central charge is the zero-mode $\cpz$ that appears in the transformation law \eqref{ifit1} of gravitational data.

Before turning to the Lie algebra of BMS$_2$, we stress again that the restriction to \textit{exact} one-forms yields a warped Virasoro subgroup of BMS$_2$. At the level of central extensions, this relation `explains' why the derivative-free cocycle \eqref{coq} has not been spotted so far in the physics literature. Indeed, writing $\eta=\di\sigma$ for some periodic function $\sigma$ reduces $\sfC(f,\eta)$ to the zero-mode of $(f^{-1})^*\sigma-\sigma$, which trivializes the cocycle $\sfC$.\footnote{For \textit{exact} BMS$_2$ translations, one can define the one-cochain $\sfK(f,\sigma)\equiv-\oint\sigma$, whereupon the extension \eqref{coq} becomes a trivial cocycle $\sfC(f_1,\di\sigma_2)=\sfK\big((f_1,\sigma_1)\cdot(f_2,\sigma_2)\big)-\sfK(f_1,\sigma_1)-\sfK(f_2,\sigma_2)$.} By contrast, the cocycle \eqref{cok} remains non-trivial in that subgroup, since it then pairs the one-cocycle $f''/f'$ with the function $\sigma$, reproducing the non-trivial extension of the warped Virasoro group investigated in \cite{Afshar:2015wjm}.

\subsection{The \texorpdfstring{BMS$_2$}{BMS2} algebra}
\label{sebalg}

In accordance with the definitions just provided, the centrally extended BMS$_2$ algebra consists of triples $(\eps,\eta,\bz)$ where $\eps=\eps(\phii)\der_{\phii}$ is a vector field (an infinitesimal diffeomorphism), $\eta=\eta(\phii)\di\phii$ is a one-form (an infinitesimal translation), and $\bz\in \RR^3$ (or $\CC^3$ if needed) is again a triple of central terms. The BMS$_2$ group acts on this space according to the adjoint representation, whose computation is omitted here for brevity and relegated instead to appendix \ref{app1}. The differential of the adjoint at the identity then yields the Lie bracket of BMS$_2$,
\be
\begin{split}
&\big[(\eps_1,\eta_1,\bz_1),(\eps_2,\eta_2,\bz_2)\big]
=\\
&=
\Big(
[\eps_1,\eps_2],(\eps_1\eta_2)'-(\eps_2\eta_1)';
\oint\frac{\di\phii}{24\pi}\,\eps_1'''\eps_2,
\oint\frac{\di\phii}{2\pi}\big(\eps_2'\eta_1-\eps_1'\eta_2\big),
\oint\frac{\di\phii}{2\pi}\big(\eps_2\eta_1-\eps_1\eta_2\big)
\Big)
\end{split}
\label{bak}
\ee
where $[\eps_1,\eps_2]\equiv \eps_1\eps_2'-\eps_2\eps_1'$ is the standard Lie bracket of vector fields on the circle, and the centreless part more generally reproduces the algebra defined in \eqref{liba} in terms of asymptotic Killing vector fields. In the central terms, the factor $\eps_1'''$ is the infinitesimal Schwarzian derivative, while the last two extensions coincide with \cite[eqs.\ (17)--(18)]{Ovsienko:1994im}.\footnote{Our notation differs from that of \cite{Ovsienko:1994im}: what we call $\eps_1,\eps_2$ and $\eta_1,\eta_2$, they respectively write as $f,g$ and $a,b$.}

Following the earlier definitions \eqref{s3q}, the bracket \eqref{bak} can also be displayed in the Fourier basis $L_m\equiv(-ie^{im\phii}\der_{\phii},0;\boldsymbol0)$, $Q_m\equiv(0,-ie^{im\phii}\di\phii;\boldsymbol0)$ with central charges $X\equiv(0,0;-i,0,0)$, $Y\equiv(0,0;0,-i,0)$ and $Z\equiv(0,0;0,0,-i)$. Eq.\ \eqref{bak} is then equivalent to the algebra initially announced in eqs.\ \eqref{t1b}--\eqref{b1b}:
\begin{align}
\label{bra1}
[L_m,L_n]
&=
(m-n)L_{m+n}
+
\frac{X}{12}m^3\,\delta_{m+n,0}\,,\\[5pt]
\label{bra2}
[L_m,Q_n]
&=
-(m+n)Q_{m+n}
+
\big(Y\,m-iZ\big)\,\delta_{m+n,0}\,,\\[5pt]
\label{bra3}
[Q_m,Q_n]
&=
0\,.
\end{align}
This extends the asymptotic symmetry algebra \eqref{libam} by a Virasoro central charge $X$ and two `exotic' central terms $Y,Z$ whose coefficients ($m$ and $1$) are strikingly different from the $m^3$ encountered in the BMS$_3$ algebra \cite{Barnich:2006av} or the $m^2$ of the warped Virasoro algebra \cite{Detournay:2012pc,Afshar:2015wjm}. In particular, $Z$ implies that rigid rotations and translations no longer commute, since $[L_0,Q_0]=-iZ$. Note again that the warped Virasoro algebra with vanishing Abelian zero-mode is a subalgebra of \eqref{bra1}--\eqref{bra3} obtained by discarding $Q_0$, defining $P_m\equiv mQ_m$ for non-zero $m$ and declaring $P_0\equiv iZ$. The central charge $Z$ then plays the role of the translational Casimir of the warped Virasoro algebra while $Y$ is the `twist' extension of Rindler holography \cite{Afshar:2015wjm}.

\subsection{Coadjoint representation and orbits of \texorpdfstring{BMS$_2$}{BMS2}}
\label{serta}

The Schwarzian description of JT gravity is closely related to coadjoint orbits of the asymptotic symmetry group, so we now describe the coadjoint representation of BMS$_2$ and classify its orbits. Owing to the gravitational transformation laws \eqref{28}, we focus on coadjoint vectors with non-zero central charges $(b,c)$ dual to $(Y,Z)$. As we shall see, the resulting BMS$_2$ orbits are `large' in that their codimension in $\text{Diff}\,S^1$ vanishes, in contrast to Virasoro orbits \cite{Witten:1987ty,Guieu}, BMS$_3$ orbits \cite{Barnich:2015uva} or warped orbits \cite{Afshar:2015wjm}. This has important consequences for the BMS-Schwarzian actions described in section \ref{seBeSch}, which turn out to lack saddle points as a result. Note, however, that we systematically focus on \textit{real} functions and central charges, in keeping with standard literature on the Virasoro group \cite{Guieu}. Complexification will only be addressed at the level of the Lie algebra, in terms of stabilizers of coadjoint orbits.

\paragraph{Coadjoint representation.} The coadjoint representation of any Lie group acts on the dual of its algebra. Elements of the BMS$_2$ algebra are triples $(\eps,\eta,\bz)$, so the (smooth) dual space consists of triples $(\cT,\cQ,\bc)$, where $\cT(\phii)$ and $\cQ(\phii)$ are functions while $\bc=(a,b,\cpz)$ is a triple of central charges. Owing to the respective conformal weights $(-1,1)$ of $(\eps,\eta)$, one should think of their duals $(\cT,\cQ)$ as densities with respective weights $(2,0)$. In other words, $\cT$ is akin to a CFT stress tensor, while $\cQ$ is a genuine function (with zero weight) to which we shall refer as a `momentum' since it is dual to `translations' $\eta$.\footnote{Incidentally, the fact that the central extension $Y$ in \eqref{bra2} is non-trivial is precisely due to the vanishing weight of the `current' $\cQ(\phii)$ under reparametrizations \cite{Daniel-Jacob}.} The pairing between the BMS$_2$ algebra and its dual is thus\footnote{In contrast to 3D gravity, the `charges' \eqref{copa} do \textit{not} coincide with surface charges of 2D gravity, as the latter are pointwise quantities on the boundary \cite{Grumiller:2013swa,Grumiller:2015vaa}. The pairing \eqref{copa} is nevertheless useful because it leads to the coadjoint representation, which coincides with the action of asymptotic symmetries on phase space.}
\be
\label{copa}
    \big<(\cT,\cQ,\bc),(\eps,\eta,\bz)\big>
    \equiv
    \frac{1}{2\pi}
    \oint\di\phii
    \big({\cT}(\phii)\varepsilon(\phii)+{\cQ}(\phii)\eta(\phii)\big)
    +ax+by+\cpz z\,.
\ee
The coadjoint representation is then defined so as to leave this pairing invariant, in the sense that $\langle\Ad^*_{(f,\eta)}(\cT,\cQ,\bc),(\eps,\eta,\bz)\rangle\equiv\langle(\cT,\cQ,\bc),\Ad_{(f,\eta)}^{-1}(\eps,\eta,\bz)\rangle$. This fixes all central charges, while $\cT$ and $\cQ$ transform non-trivially according to $\Ad^*_{(f,\eta)}(\cT,\cQ,\bc)\equiv\big(\widetilde\cT,\widetilde\cQ,\bc\big)$. Since the weights of $\cT$ and $\cQ$ are known, the only problem is to find the inhomogeneous terms of these transformations. The solution follows from the adjoint representation worked out in appendix \ref{app1}, and is written in eqs.\ \eqref{toff}--\eqref{boff} below. Including a change of argument for readability, one thus finds
\begin{align}
\label{ttj}
\,\widetilde\cT(f(\phii))
&=
\frac{1}{f'^2(\phii)}\Big[%
\cT
+\frac{a}{12}\sfS[f]
-b \big(f'\eta\circ f\big)'
+(\cQ'+\cpz)f'\eta\circ f
\Big]\,,\\
\label{ttq}
\widetilde\cQ(f(\phii))
&=
\cQ(\phii)-b\log f'(\phii)-\cpz(f(\phii)-\phii)\,,\Big.
\end{align}
where all terms on the right-hand side are implicitly evaluated at $\phii$ and $\sfS[f]\equiv f'''/f'-(3/2)(f''/f')^2$ is the usual Schwarzian derivative. These equations are the BMS$_2$ analogue of the transformation law of a (warped) CFT stress tensor \cite{Detournay:2012pc,Afshar:2015wjm}. In particular, the first terms on the right-hand side of \eqref{ttj} reproduce the standard CFT transformation of $\cT$, while the other terms are new. The transformation law \eqref{ttq} of $\cQ$ is especially unusual, as it involves the exotic cocycles $\log f'$ and $(f(\phii)-\phii)$. We will soon see that the term $(f(\phii)-\phii)$ has dramatic consequences for coadjoint orbits and Schwarzian actions.

The coadjoint representation of the BMS$_2$ algebra is the infinitesimal form of \eqref{ttj}--\eqref{ttq}, obtained by writing $f(\phii)\sim\phii+\eps(\phii)$ and expanding to first order in $\eps,\eta$. The result reads
\be
\label{itra}
\delta_{\eps,\eta}\cT
=
\eps\cT'+2\eps'\cT-\frac{a}{12}\eps'''+b\eta'-\eta(\cQ'+\cpz)\,,
\qquad
\delta_{\eps,\eta}\cQ
=
b\eps'
+\eps(\cQ'+\cpz)\,,
\ee
which coincides with the gravitational transformation laws \eqref{ifit1} for $a=0$ and $b=-1$. This includes, as a special case, the coadjoint representation of the warped Virasoro algebra with vanishing level \cite{Afshar:2015wjm}. Indeed, letting $\cP\equiv\cpz+\cQ'$ and $\eta\equiv\sigma'$ for some function $\sigma$, the infinitesimal transformations \eqref{itra} become
\be
\delta\cT
=
\eps\cT'+2\eps'\cT-\frac{a}{12}\eps'''-\sigma'\cP+b\sigma''\,,
\qquad
\delta\cP
=
\eps\cP'+b\eps''\,,
\ee
which coincides with \cite[eq.\ (3.12)]{Afshar:2015wjm} up to minor differences in notation. Similarly, one verifies that eqs.\ \eqref{ttj}--\eqref{ttq} reduce to the coadjoint representation of the warped Virasoro group (in terms of $\cT$ and $\cP=\cpz+\cQ'$) when $\eta$ is exact: see eqs.\ (A.21)--(A.22) of \cite{Afshar:2015wjm}. This is consistent with the fact, emphasized throughout, that the warped Virasoro group is a subgroup of BMS$_2$.

\paragraph{A single coadjoint orbit.} The coadjoint representation \eqref{ttj}--\eqref{ttq} displays a common feature of all semi-direct products with an Abelian factor \cite{Rawnsley1975}: the `momentum' transformation law \eqref{ttq} is insensitive to translations $\eta$ and involves no stress tensor $\cT$. (Indeed, the same general structure appears in the Poincar\'e group, the BMS$_3$ group \cite{Barnich:2015uva} and the warped Virasoro group with vanishing level \cite{Afshar:2015wjm}.) As a result, any coadjoint orbit of BMS$_2$ is a fibre bundle over the orbit of some momentum $\cQ$ under diffeomorphisms. We therefore focus for now on eq.\ \eqref{ttq} alone, and classify the orbits of momenta under reparametrizations.

What is special about BMS$_2$ (in contrast to the other examples mentioned above) is that there is only \textit{one} momentum orbit at fixed central charges---at least provided $b\neq0$ and $\cpz\neq0$ are both real. In fact, the following statement holds:
\begin{remark}
\label{thm}
The coadjoint representation of the real BMS$_2$ group with real non-zero central charges $b,c$ can be used to map any momentum $\cQ$ on $\widetilde\cQ=0$. The orbit of $\cQ=0$ under diffeomorphisms is thus (diffeomorphic to) the entire group $\text{Diff}\,S^1$.
\end{remark}
The proof is straightforward if $b\neq0$ and $\cpz\neq0$ are both real: given $\cQ(\phii)$, set $\widetilde Q=0$ in \eqref{ttq} to find an ordinary differential equation for $f(\phii)$ whose general solution is\footnote{We write the solution as $\hat f$ instead of $f$ to stress that it is a specific group element, determined (uniquely) by $\cQ$; this notation will be useful in section \ref{secomp} below.\label{footex}}
\be
\label{unimap}
\hat f(\phii)
=
\frac{b}{\cpz}
\log\left[%
K
+
\frac{\cpz}{b}\int_0^{\phii}\di\theta\,e^{(\cQ(\theta)+\cpz\,\theta)/b}
\right]
\ee
for some integration constant $K$. The latter is uniquely fixed by the requirement that $\hat f$ be a diffeomorphism of the circle ($\hat f(\phii+2\pi)=\hat f(\phii)+2\pi$), which thus yields an explicit diffeomorphism mapping $\cQ$ on $\widetilde\cQ=0$.\footnote{When $b=0$, one can similarly write $\hat f(\phii)=\phii+\cQ(\phii)/\cpz$ to map $\cQ$ on $\widetilde\cQ=0$, but this generally fails to satisfy the condition $\hat f'>0$ if $\cQ(\phii)$ happens to be `too steep'; we therefore assume $b\neq0$ from now on.} \hfill$\blacksquare$

A corollary of theorem \ref{thm} is that any zero-mode configuration $\cQ$ can be mapped on any other one. This is obvious from eq.\ \eqref{ttq}: the rotation $f(\phii)=\phii+\theta$ sends $\cQ=\text{cst}$ on $\tilde\cQ=\cQ-\cpz\,\theta$. To the extent that `rotations' in $\phii$ are really (Euclidean) time translations in JT gravity, this states that the zero-mode of $\cQ$ `rotates' at a velocity set by the central charge $\cpz$. One can also confirm these results from an analysis of stabilizers at the level of the Lie algebra: when $\cpz\neq0$ and $b$ are real, setting $\delta\cQ=0$ in the second equation of \eqref{itra} yields an equation for $\eps$ whose only $2\pi$-periodic solution is $\eps=0$, regardless of the form of $\cQ$.

Starting from the trivial `classification' just outlined, it is immediate to derive the structure of coadjoint orbits of the real BMS$_2$ group. Indeed, as explained \eg in \cite{Barnich:2015uva}, coadjoint orbits of semi-direct products are bundles of little group orbits over the cotangent bundle of a momentum orbit. In the case at hand, we have just seen that the stabilizer of any momentum (with $\cpz\neq0$) is trivial, so all little groups are trivial and any stress tensor $\cT$ can be mapped on any other one, say $\widetilde\cT=0$, by the action of suitable BMS$_2$ group elements. It follows that
\begin{remark}
\label{thm2}
The BMS$_2$ group has a single coadjoint orbit for all real values of its three real central charges such that $b\neq0$ and $\cpz\neq0$; this orbit is diffeomorphic to the cotangent bundle $T^*\text{Diff}\,S^1$.
\end{remark}
The `abstract nonsense' proof is immediate based on theorem \ref{thm} and \cite[sec.\ 2]{Barnich:2015uva}, but it is also illuminating to carry it out directly from the transformation law \eqref{ttj}. Indeed, one can use the diffeomorphism \eqref{unimap} to map $\cQ$ on $\hat f\cdot Q\equiv\widetilde\cQ=0$, thereby mapping $\cT$ on some new stress tensor $\widetilde\cT$ according to \eqref{ttj} with $\eta=0$; this is a standard CFT transformation law with central charge $a$:
\be
\label{extrabis}
\widetilde T
=
[(\hat f^{-1})']^2\,T\circ \hat f^{-1}
-\frac{a}{12}\sfS[\hat f^{-1}]\,.
\ee
As a second step, one can implement a pure translation $\hat\eta$ (letting $f=\mathbb{I}$ be the identity) to leave $\widetilde\cQ=0$ invariant while also mapping $\widetilde\cT$ on $\widetilde{\widetilde\cT}=0$;\footnote{As in theorem \ref{thm} and footnote \ref{footex}, the hat in $\hat\eta$ stresses that $\hat\eta$ is a specific translation determined by $(\cT,\cQ)$ rather than an arbitrary one.} this is the case provided $\hat\eta$ is the unique $2\pi$-periodic solution of
\be
\widetilde\cT-b\,\hat\eta'+\cpz\,\hat\eta=0\,,
\label{extra}
\ee
which will turn out to be useful in section \ref{secomp}. The composition $(\mathbb{I},\hat\eta)\circ(\hat f,0)=(\hat f,\hat\eta)$ is thus the unique BMS$_2$ group element mapping $(\cT,\cQ)$ on $(0,0)$, as was to be shown.

Again, one can confirm this result with the Lie-algebraic formulas \eqref{itra}: simultaneously setting $\delta\cT=0$ and $\delta\cQ=0$ under the assumption that $\cpz\neq0$ and $b$ are real yields $\eps=\eta=0$ as the only periodic solution, confirming that the stabilizer of any pair $(\cT,\cQ)$ is trivial. \hfill$\blacksquare$

We stress that these results hinge on the assumption that functions and central charges are all \textit{real}. The presence of complex quantities, such as the purely imaginary boundary graviton $\cP$ in eq.\ \eqref{214}, would alter our conclusions drastically. For example, setting $\delta\cQ=0$ in eq.\ \eqref{itra} with a purely \textit{imaginary} ratio $c/b$ leaves room for complex periodic solutions provided $c/b\in i\ZZ^*$. This is the situation anticipated around eq.\ \eqref{s66b}, where the stabilizer of the background $(\cT,\cP)$ was interpreted as a Euclidean isometry group. It goes without saying that the corresponding orbits are different from those described in theorems \ref{thm}--\ref{thm2}. However, we refrain from attempting to classify such complex orbits, as the complexification of diffeomorphism groups is a notoriously thorny issue \cite{Guieu}. Infinitesimal considerations of the kind just described will suffice nevertheless, since path integral computations in section \ref{secomp} will eventually localize to the identity.

\subsection{BMS-Schwarzian actions}
\label{seBeSch}

We saw in eq.\ \eqref{Sell} that the boundary action of flat JT gravity can be seen as a functional of a pair $(f,\eta)$ belonging to the BMS$_2$ group, with transformation laws \eqref{s4b} that stem from left-invariant vector fields \eqref{ss8b} on the group manifold. From that perspective, the pair $(\cT,\cQ)$ is merely a parameter specifying the background around which dilaton fluctuations are being considered. In the analogous case of JT gravity on AdS$_2$, it is then typically possible to interpret the boundary action as a `Schwarzian action', \ie as a rotation generator on a Virasoro coadjoint orbit \cite{Stanford:2017thb}. Is there a similar interpretation in flat JT?

To begin, note that the answer is very nearly positive. Indeed, the generator of rotations $\phii\mapsto\phii+\theta$ (\ie Euclidean time translations) is the zero-mode of the `stress tensor' $\cT(\phii)$. More precisely, on the BMS$_2$ orbit of a `seed' $(\cT,\cQ)$, the stress tensor is given by \eqref{ttj} and its zero-mode is the \textit{BMS-Schwarzian action}
\be
\label{s13t}
S_{\text{BMS}}[f,\eta]
\equiv
\frac{1}{2\pi}\oint\frac{\di\phii}{f'}
\Big(%
\cT+\frac{a}{12}\sfS[f]-bf''\eta\circ f+(\cQ'+\cpz)f'\eta\circ f
\Big)\,.
\ee
This almost coincides with the JT action \eqref{Sell} when $a=0$ and $b=-1$: the only mismatch is the absence of the zero-mode of $\eta$ that appears in \eqref{Sell}. The difference was to be expected, since the would-be equations of motion derived by varying $\eta\circ f$ in \eqref{s13t} read $-bf''+(\cQ'+c)f'=0$, and only admit solutions that satisfy $f'>0$ and $f(\phii+2\pi)=f(\phii)+2\pi$ in the exceptional case $\cpz=0$. In other words, the action \eqref{s13t} generally has no saddle points at all, in stark contrast to the JT action \eqref{Sell} that was precisely designed so as to admit saddles.

The existence of saddle points thus hinges on the additional term $\propto\cpz\oint\eta$ on the far right-hand side of \eqref{Sell}. One way to justify its presence, irrespective of flat JT dynamics, is to investigate the transformation law of the Schwarzian action \eqref{s13t} under left- and right-invariant vector fields. Indeed, recall from eq.\ \eqref{ss8b} that left-invariant vector fields act on pairs $(f,\eta)$ so as to reproduce the transformation of the dilaton under asymptotic symmetries; one then finds that the variation of the Schwarzian action under a \textit{left} pure translation $\zeta$ is
\be
\label{dess}
\delta^{\text{L}}_{(0,\zeta)}S_{\text{BMS}}
=
\frac{1}{2\pi}
\oint\frac{\di\phii}{f'}
\Big(%
-b\,\frac{f''}{f'}+\cQ'+\cpz\Big)\zeta
\,,
\ee
where it is understood that $\zeta$ acts on $(f,\eta)$ according to eq.\ \eqref{ss8b}, while $(\cT,\cQ)$ are fixed, and the integrand is evaluated at $\phii$. A similar computation can be carried out for \textit{right}-invariant vector fields whose action on $(f,\eta)$ is obtained similarly to the left-invariant equation \eqref{ss8b}:
\be
\label{saab}
\big(\delta_{(\eps,\zeta)}^{\text{R}}f,\delta_{(\eps,\zeta)}^{\text{R}}\eta\big)
\equiv
\frac{\partial}{\partial t}\bigg|_{t=0}\Big[(e^{t\eps},t\zeta)\cdot(f,\eta)\Big]
\stackrel{\text{\eqref{mu}}}{=}
\big(\eps\circ f,\zeta-(\eps\eta)'\big)\,.
\ee
Applying this to the Schwarzian functional \eqref{s13t} with $\eps=0$ yields
\be
\label{dett}
\delta^{\text{R}}_{(0,\zeta)}S_{\text{BMS}}
=
\frac{1}{2\pi}
\oint\di\phii
\Big(
-b\,\frac{f''}{f'}+\cQ'+\cpz
\Big)\zeta\circ f\,,
\ee
which only differs from \eqref{dess} by the replacement of $\zeta$ by $\zeta\circ f$ and the crucial absence of the denominator $f'(\phii)$. Indeed, let us compare these transformation laws to those of the zero-mode of $\eta$: using either the left action \eqref{ss8b} or the right action \eqref{saab}, one finds
\be
\label{deta}
\delta^{\text{L,R}}_{(\eps,\zeta)}\oint\eta
=
\oint\zeta
\ee
for any $(\eps,\zeta)$ in the BMS$_2$ algebra. Comparing this with the transformations \eqref{dess}--\eqref{dett}, it is clear that neither the Schwarzian action \eqref{s13t} nor the full boundary action \eqref{Sell} is invariant under all pure translations---regardless of whether one acts with left- or right-invariant vector fields. But an exception occurs for \textit{constant} translations ($\zeta(\phii)=\text{cst}\equiv\zeta_0$), in which case
\be
\delta^{\text{L}}_{(0,\zeta_0)}S_{\text{BMS}}
=
\frac{\zeta_0}{2\pi}
\oint\frac{\di\phii}{f'(\phii)}(\cQ'+\cpz)
\qquad\text{and}\qquad
\delta^{\text{R}}_{(0,\zeta_0)}S_{\text{BMS}}
=
\cpz\,
\zeta_0\,.
\ee
It then follows from \eqref{deta} that the full boundary action $2\pi S_{\text{BMS}}-c\oint\eta$ is invariant under constant translations generated by \textit{right}-invariant vector fields. This is the sense in which adding the zero-mode of $\eta$ to the Schwarzian \eqref{s13t} enhances the symmetries of the theory.

\section{Thermal partition function}
\label{sePAF}

Here we evaluate the partition function of flat JT gravity at finite temperature, as described by the action \eqref{Sell} with $\ell=\infty$. The plan is as follows: we first summarize the problem at hand, then choose a suitable measure on the BMS$_2$ group manifold, and finally integrate the exponential of the action to obtain the partition function. The result will be consistent with earlier similar considerations in the literature \cite{Afshar:2019tvp}; it is one-loop exact and exhibits the expected symmetry enhancement that occurs on backgrounds that cover Euclidean space.

\subsection{Setting the stage}

The difference between the flat JT action \eqref{Sell} and the Schwarzian \eqref{s13t} implies that the former is \textit{not} a Schwarzian: it is not a Hamiltonian function generating a U(1) action on a coadjoint orbit endowed with its Kirillov-Kostant symplectic form \cite{KirillovLectures}. One may therefore worry that the partition function of the theory cannot be evaluated exactly, in contrast to more standard cases where the Duistermaat-Heckman theorem localizes the path integral \cite{Stanford:2017thb,Afshar:2019tvp,Godet:2020xpk}. (See \eg \cite{Cremonesi:2013twh} for a pedagogical introduction to localization.) Nevertheless, one-loop exactness is actually trivial in the present case, as the dilaton component $\eta$ acts as a Lagrange multiplier that enforces the equation of motion for $f$, somewhat similarly to BMS$_3$ characters \cite{Oblak:2015sea}. At a deeper level, a conceivable explanation of this simple result is that the flat JT action \eqref{Sell} is a U(1) generator on some larger symplectic manifold (say the cotangent bundle of BMS$_2$). We refrain from attempting to build such a framework, as it will be unnecessary anyway.

Let us set the stage for the problem we wish to address. Fixing a pair $(\cT,\cQ)$ that specifies some gravitational background \eqref{214} with $\cP=\cpz+\cQ'$, our goal is to evaluate a dilatonic partition function
\be
\label{z}
Z
\equiv
\int\limits_{\text{BMS}_2}\cD f\,\cD\eta\,e^{-S[f,\eta]}
\ee
where the path integral is taken over all elements $(f,\eta)$ of the (centreless) BMS$_2$ group and the Euclidean action functional generalizes the flat limit of the JT boundary action \eqref{Sell}:
\be
\label{sefe}
S[f,\eta]
=
-\frac{i\kappa\Lambda}{\cpz}
\oint\frac{\di\phii}{f'(\phii)}
\bigg[%
\cT+\frac{a}{12}\sfS[f]
+\Big(\big(\cpz+\cQ'(\phii)\big)\,f'-bf''-\cpz f'^2\Big)\eta\circ f
\bigg].
\ee
The integration measure $\cD f\,\cD\eta$ remains to be fixed; following the literature on Schwarzian actions and 2D gravity (see \eg \cite{Stanford:2017thb,Afshar:2019tvp}), we shall choose it to be (right-)invariant under the asymptotic symmetry group---BMS$_2$ in the case at hand. Note again that the prefactor $-i\kappa\Lambda/\cpz\stackrel{\text{\eqref{s6b}}}{=}2\pi i\kappa/\cA$ in \eqref{sefe} is real when $\cpz$ is purely imaginary. This will indeed be assumed throughout, and it is consistent with the inclusion of the Euclidean plane in the phase space of Euclidean metrics \eqref{214}.

The expression \eqref{sefe} mimics the BMS-Schwarzian \eqref{s13t}, including all three BMS$_2$ central charges, but it crucially also includes the vacuum energy zero-mode $\propto\oint\eta$ that ensures the existence of saddle points, as in the gravitational action \eqref{Sell}. Accordingly, it makes sense to evaluate the path integral \eqref{z} perturbatively around saddle points; this will yield an exact result thanks to the flatness of the measure along $\eta\circ f$. To carry out the integration, we start by building a right-invariant measure (section \ref{sehaar}), then exploit it to evaluate the partition function (section \ref{secomp}).

\subsection{An invariant measure on \texorpdfstring{BMS$_2$}{BMS2}}
\label{sehaar}

As emphasized above, typical partition functions of the form \eqref{z} in JT gravity are integrals over coadjoint orbits of asymptotic symmetry groups \cite{Stanford:2017thb}. The (Liouville) path integration measure is then induced by the Kirillov-Kostant symplectic form on an orbit. In the case at hand, theorem \ref{thm2} ensures that the entire BMS$_2$ group is itself (diffeomorphic to) a coadjoint orbit. The Liouville measure thus coincides with the Haar measure on the group manifold, and can be found by evaluating expectation values of the (left or right) Maurer-Cartan form. We now summarize this computation, while details involving coadjoint orbits and central charges are relegated to appendix \ref{appE}.

It is worth recalling first how the Haar measure is built for any Lie group. Let $G$ be an $n$-dimensional group with Lie algebra $\mathfrak{g}$, and let $\mu_0$ be a (translation-invariant) volume form on $\mathfrak{g}$; this form is unique up to normalization. A right-invariant measure $\mu$ on $G$ is then defined by the right Maurer-Cartan form of $G$,
\be
\mu(v_1,...,v_n)
\equiv
\mu_0\big(\di(R_{g^{-1}})_gv_1,...,\di(R_{g^{-1}})_gv_n\big)
\label{lemm}
\ee
where $g$ is any point on $G$, $R_g$ means right multiplication by $g$, and the $v_i$'s are tangent vectors of $G$ at $g$. The application to BMS$_2$ is straightforward: since the Lie algebra consists of pairs $(\eps,\eta)$, the only translation-invariant measure in $(\eps,\eta)$ space is $\mu_0\propto\prod_{\phii\in[0,2\pi)}\delta\eps(\phii)\wedge\delta\eta(\phii)$, where $\delta$ denotes a `vertical' functional exterior derivative. The invariant measure \eqref{lemm} then takes the same form, except that $\delta\eps$ and $\delta\eta$ are replaced by the corresponding components of the right Maurer-Cartan form. The latter is the Lie algebra-valued one-form
\be
\label{e44}
\delta\big(R_{(f,\eta)^{-1}}\big)_{(f,\eta)}
=
\Big(%
\delta f\circ f^{-1},
\delta\eta
+\big(\eta\,\delta f\circ f^{-1}\big)'
\Big)\,,
\ee
obtained \eg by using the centreless group operation \eqref{mu} to compute the right logarithmic time derivative $\partial_{\tau}[(f_{\tau},\eta_{\tau})(f_t,\eta_t)^{-1}]$ of a path $(f_t,\eta_t)$ in BMS$_2$ (see appendix \ref{appE}). It follows that the right Haar measure on the centreless BMS$_2$ group, evaluated at the point $(f,\eta)$, reads
\be
\label{omega}
\text{Haar measure}
=
N\prod_{\phii\in[0,2\pi)}\delta f\big(f^{-1}(\phii)\big)\wedge\delta\eta(\phii)
=
N\prod_{\phii\in[0,2\pi)}\delta f(\phii)\wedge\delta(\eta\circ f)(\phii)
\ee
for some unimportant normalization $N$. Note that the extra term $\sim(\eta\,\delta f)'$ in the Maurer-Cartan form \eqref{e44} drops out thanks to the antisymmetric wedge product of one-forms. Incidentally, it follows that the right measure coincides with the left one, as in the standard Virasoro group \cite{Stanford:2017thb}.

The measure \eqref{omega} is flat on the space of functions $(f,\eta)$, suggesting that the partition function \eqref{z} may be computed with straightforward path integral methods. There is a problem, however: the action of BMS$_2$ on phase space is projective, including non-commuting zero-modes $L_0$, $Q_0$ (recall the algebra \eqref{bra1}--\eqref{bra3}). As a result, the correct invariant measure is not quite the one written here, but its centrally extended generalization on a coadjoint orbit, crucially sensitive to the central charge-dependent complex stabilizers mentioned at the very end of section \ref{serta}. The relevant computation involves the Kirillov-Kostant symplectic form $\Omega$ and is presented in appendix \ref{appE} for brevity. In particular, the symplectic structure is displayed in eq.\ \eqref{symplecticformBis}; it leads to a Liouville measure $\cD f\,\cD\eta=\Omega^{\wedge\infty}$ that takes the general form
\be
\label{lime}
\text{Liouville measure}
=
N\,\,
\text{Pf}[\Omega](f)
\prod_{\phii\in[0,2\pi)}
\delta f(\phii)\wedge\delta(\eta\circ f)(\phii)
\equiv
\cD f\,\cD\eta\,.
\ee
Here $\text{Pf}[\Omega](f)$ is the Pfaffian of the symplectic form evaluated at $f$, and it \textit{only} depends on $f$. Accordingly, the measure \eqref{lime} is \textit{flat} in $\eta\circ f$, even after including central charges. The action \eqref{sefe} thus ensures that the integral over $\eta\circ f$ yields a straightforward Dirac distribution on the equation of motion of $f$. Localization is therefore trivial, as in \cite{Oblak:2015sea}.

\subsection{Computation of the path integral}
\label{secomp}

Here we compute the one-loop exact partition function \eqref{z}. This is done for any background $(\cT,\cQ)$ by first making the dependence on $(\cT,\cQ)$ explicit and reducing the problem to that of a `reference' background $(0,0)$. The actual path integration is then carried out as a second step, by evaluating the Pfaffian in \eqref{lime} at the saddle point and multiplying it by the appropriate determinant.

\paragraph{Background dependence.} If the action $S$ in \eqref{z} were a pure Schwarzian, right-invariance of the measure \eqref{lime} would immediately imply that the value of the integral \eqref{z} is independent of the orbit representative $(\cT,\cQ)$ chosen as `background', and only depends on its orbit. One would then conclude from theorem \ref{thm2} that the choice $(\cT,\cQ)=(0,0)$ entails no loss of generality, since BMS$_2$ has a unique orbit. But there is a catch: the action \eqref{sefe} is a Schwarzian \textit{plus} a zero-mode of $\eta$, so right-invariance of the measure does \textit{not} imply that the integral \eqref{z} only depends on the orbit of $(\cT,\cQ)$.

In fact, it is a simple matter to exhibit the dependence of \eqref{z} on $(\cT,\cQ)$: for generic central charges, theorem \ref{thm2} ensures that there exists a unique BMS$_2$ group element $(g,\zeta)$ such that $(\cT,\cQ,\bc)=\Ad^*_{(g,\zeta)}(0,0,\bc)$ in terms of the coadjoint representation of section \ref{serta}. This group element is the inverse of the one built in the proof of theorems \ref{thm}--\ref{thm2}, \ie $g=\hat f^{-1}$ is the inverse of the diffeomorphism \eqref{unimap} determined by $\cQ$, while $\zeta=-\hat f^*\hat\eta$ involves the unique solution $\hat\eta$ of \eqref{extra}, determined by $\cQ$ and $\cT$. This can be plugged in the action \eqref{sefe} to yield
\be
\label{s18b}
S_{(\cT,\cQ)}[f,\eta]
=
S_{(0,0)}\Big[(f,\eta)\cdot\big(\hat f^{-1},-\hat f^*\hat\eta\big)\Big]
+
i\kappa\Lambda\,\oint\di\phii\,\hat\eta(\phii)\,,
\ee
where the subscript of $S$ indicates the corresponding background and the argument of $S$ on the right-hand side involves the BMS$_2$ group operation \eqref{mu}; this shift of argument will eventually be unimportant thanks to the right-invariance of the measure. The second term on the right-hand side of \eqref{s18b} is a non-local functional of $(\cT,\cQ)$ that can be found by integrating eq.\ \eqref{extra} over the circle and letting $\widetilde\cT$ be the conformally transformed stress tensor \eqref{extrabis} to obtain
\be
\label{t19b}
\cpz
\,
E[\cT,\cQ]
\equiv
-\cpz
\,\oint\di\phii\,\hat\eta(\phii)
=
\oint\frac{\di\phii}{\hat f'(\phii)}\Big[\cT(\phii)+\frac{a}{12}\sfS[\hat f](\phii)\Big],
\ee
where the diffeomorphism $\hat f(\phii)$ takes the form \eqref{unimap} in terms of $\cQ$, and the notation `$E$' is adopted to stress that the zero-mode of $\hat\eta(\phii)$ should now be interpreted as a `ground state energy'. The key point is that \eqref{t19b} depends neither on $f$ nor on $\eta$ (the dynamical fields), so it can be pulled out of the partition function \eqref{z}. Since the latter involves a right-invariant measure, it can be recast as
\be
Z
=
e^{i\kappa\Lambda\,E[\cT,\cQ]}
\int\limits_{\text{BMS}_2}\cD f\,\cD\eta\,e^{-S_{(0,0)}[f,\eta]}\,,
\label{b19b}
\ee
with an integral over $(f,\eta)$ that now only involves the action \eqref{sefe} on the background $(\cT,\cQ)=(0,0)$. This is the sense in which theorem \ref{thm2} allows us to focus on the orbit representative $(\cT,\cQ)=(0,0)$ without loss of generality.

Note that the prefactor $\sim e^{\#\,E}$ of \eqref{b19b} is nearly trivial in many cases of physical interest. For instance, if the pair $(\cT,\cQ)$ has $\cQ(\phii)=\text{cst}$, then the uniformizing diffeomorphism \eqref{unimap} is just a rotation and \eqref{t19b} yields $\cpz\,E=2\pi\,\cT_0$ in terms of the zero-mode of $\cT(\phii)$. Complications only occur when $\cQ'\neq0$, in which case the uniformizing map \eqref{unimap} is more involved and $E$ becomes the zero-mode \eqref{t19b} of a conformally transformed stress tensor with $\hat f'\neq1$.

\paragraph{Path integration.} Let us now compute the path integral in eq.\ \eqref{b19b}. Taking $(\cT,\cQ)=(0,0)$ in the action \eqref{sefe} yields
\be
\int
\cD f\,\cD\eta\,e^{-S_{(0,0)}[f,\eta]}
=
\int
\cD f\,\cD\eta\,
\exp\bigg[
\frac{i\kappa\Lambda}{\cpz}
\oint\frac{\di\phii}{f'(\phii)}
\Big(%
\frac{a}{12}\sfS[f]
-bf''\eta\circ f
+\cpz(f'-f'^2)\eta\circ f\Big)
\bigg]
\ee
where $\cD f\,\cD\eta$ is the Liouville measure \eqref{lime}. The equations of motion simply set $f'=1$ and $\eta=0$, so we expand the action to second order around its saddle point by letting $f(\phii)=\phii+\eps(\phii)$ and assuming that $\eps$ and $\eta$ are of the same order. This gives
\be
Z
=
N\,
e^{i\kappa\Lambda\,E[\cT,\cQ]}\,
\text{Pf}[\Omega](\mathbb{I})
\int\cD\eps\,\cD\eta\,
\exp\bigg[
\frac{i\kappa\Lambda}{\cpz}
\oint\di\phii
\Big(%
\frac{a}{24}\eps''^2
-b\eps''\eta
-\cpz\eps'\eta\Big)
\bigg]
\ee
where $\text{Pf}[\Omega](\mathbb{I})$ is the Pfaffian of the symplectic form \eqref{symplecticformBis} evaluated at the identity in BMS$_2$, while the measures $\cD\eps$ and $\cD\eta$ are now standard `flat' functional measures. To proceed, we assume that $\eps$ is real while $\eta$ is purely imaginary, and expand them in Fourier modes. (As before, $\cpz$ is assumed to be purely imaginary while $\Lambda$ is real.) Performing the resulting Gaussian/Fresnel integrals eventually yields
\be
\label{zprod}
Z
=
N\,
e^{i\kappa\Lambda\,E[\cT,\cQ]}\,
\text{Pf}[\Omega](\mathbb{I})
\prod_{m=1}^{+\infty}\frac{-\cpz^2}{4\kappa^2\Lambda^2m^2}
\prod_{m=1}^{+\infty}\frac{1}{\cpz^2+b^2m^2}\,.
\ee
Note the divergence of the $n^{\text{th}}$ factor of the infinite product when $c=\pm ibn$. We return to this exceptional case below, assuming for the time being that $\cpz$ is `generic' in the sense that it cannot be written as $\pm ibn$ for any integer $n$.

It now only remains to evaluate the Pfaffian term in \eqref{zprod}, which involves once more some symplectic considerations that are relegated to appendix \ref{appE}. In terms of Fourier modes $\eps_n$, $\eta_n$ of $\eps$ and $\eta$ respectively, one thus finds from eq.\ \eqref{symplecticformBis} that the symplectic form at the identity is
\be
\label{symplecticform2}
  \Omega
  =
  \sum_{m=1}^{+\infty}\bigg[%
  (...)\delta\eps_m\wedge\delta\eps_{-m}
  -(\cpz+ibm)\delta\eta_m\wedge\delta\eps_{-m}
  -(\cpz-ibm)\delta\eta_{-m}\wedge\delta\eps_m
  +\cpz\,\delta\eps_0\wedge\delta\eta_0\bigg]
\ee
where the omitted coefficient (...) denotes an irrelevant $m$-dependent term.\footnote{This coefficient also depends on the Virasoro central charge $a$ and would play a key role if we were integrating over a \textit{Virasoro} coadjoint orbit \cite{Stanford:2017thb}, but this is not so in the BMS$_2$ case treated here.} The $c\pm ibm$ coefficients may be seen as structure constants appearing in the BMS$_2$ Lie bracket \eqref{bra2}. The Pfaffian of \eqref{symplecticform2} therefore reads
\be
\label{pfaffian}
\text{Pf}[\Omega](\mathbb{I})
=
\cpz\prod_{m=1}^{+\infty}(\cpz^2+b^2m^2)\,,
\ee
where we note again the cancellation that occurs when $\cpz=\pm ibn$ for some integer $n$. Plugging this back in the partition function \eqref{zprod} and assuming once more that $\cpz$ is generic cancels the one-loop determinant $\prod\frac{1}{c^2+b^2m^2}$, so one finally finds
\be
\label{zz}
Z
=
N\,
\cpz\,
e^{i\kappa\Lambda\,E[\cT,\cQ]}\,
\prod_{m=1}^{+\infty}\frac{-\cpz^2}{4\kappa^2\Lambda^2m^2}
\propto
e^{i\kappa\Lambda\,E[\cT,\cQ]}\,
\prod_{m=1}^{+\infty}\frac{1}{m^2}\,.
\ee
On the far right-hand side, the neglected proportionality factor only contains temperature-independent (and possibly divergent) contributions that are ultimately irrelevant for thermodynamics. What is crucial is the product of $1/m^2$ factors: had we worked in terms of Matsubara frequencies instead of Fourier modes, each such factor would have taken the form $\beta^2/(2\pi m)^2$. Zeta function regularization can then be used to write
\be
\prod_{m=1}^{+\infty}\Big(\frac{\beta}{2\pi m}\Big)^2
=
\exp\big[2\zeta'(0)+2\log(\tfrac{\beta}{2\pi})\,\zeta(0)\big]
=
\frac{1}{\beta}\,,
\label{tt20b}
\ee
implying that the partition function \eqref{zz} satisfies
\be
\label{zfinal}
Z
\propto
e^{i\kappa\Lambda\,E[\cT,\cQ]}\,
\frac{1}{\beta}
\ee
up to temperature-independent factors.  This result was to be expected: the action \eqref{sefe} is invariant under Euclidean time translations and rigid translations, so the modes $\eps_0$ and $\eta_0$ are modded out when the path integral \eqref{z} is performed on generic gravitational backgrounds, as was assumed here. The result \eqref{zfinal} thus reproduces similar one-loop partition functions found in the warped Schwarzian theory \cite{Afshar:2019tvp}.

The thermal partition function on $n$-fold covers of the Euclidean plane differs from \eqref{zfinal}. Indeed, in that case one has $\cpz=\pm ibn$ for some integer $n$, and the symplectic form \eqref{symplecticform2} acquires one additional (complex) degenerate direction. One should therefore mod out this one extra dimension when performing the integral, resulting in an infinite product $Z_{\text{$n$-fold plane}}\supset\prod_{\substack{m\in\ZZ^*,\\m\neq n}}\frac{1}{|m|}$ instead of \eqref{zz}. Reinstating Matsubara frequencies and using once more zeta function regularization, as in eqs.\ \eqref{tt20b}, yields a partition function
\be
\label{zplane}
Z_{\text{$n$-fold plane}}
\propto
e^{i\kappa\Lambda\,E[\cT,\cQ]}\,
\frac{1}{\beta^2}\,.
\ee
Here the tree-level piece takes the same form as in \eqref{zfinal} despite the presence of an additional dimension in the stabilizer; this is because the additional Fourier modes allowed by eq.\ \eqref{extra} when $\cpz=\pm ibn$ do not affect the \textit{zero-mode} of its solution $\zeta$. But the one-loop determinant $\propto\beta^{-2}$ manifestly differs from the $\beta^{-1}$ of eq.\ \eqref{zfinal}. The distinction arises from the symmetry enhancement that occurs for (covers of) the Euclidean plane as opposed to more generic backgrounds: recall our discussion of isometries at the end of section \ref{sephatemp}, as well as the complexified orbit stabilizers of section \ref{serta}. The same type of enhancement occurs in the Schwarzian theory \cite{Stanford:2017thb}, except that the relevant stabilizers in that case are those of Virasoro orbits (\ie isometries of AdS$_2$ backgrounds).

As a concluding aside, note that we could have kept track of all numerical factors in the derivation above: eqs.\ \eqref{zfinal} and \eqref{zplane} would then read
\be
Z
=
\cN\,
e^{i\kappa\Lambda\,E[\cT,\cQ]}\,
\frac{1}{\beta}\,,
\qquad
Z_{\text{$n$-fold plane}}
=
\cN\,
e^{i\kappa\Lambda\,E[\cT,\cQ]}\,
\frac{4\pi\kappa\Lambda}{b}\,
\frac{1}{\beta^{2}}\,,
\ee
where $\cN$ denotes the same normalization constant in both cases. This is likely to be useful when using a replica trick to compute thermodynamic observables, in the spirit of \cite{Engelhardt:2020qpv}, though we refrain from pursuing this line of thought here.

\section{Conclusion and outlook}

This work was devoted to the Bondi-Metzner-Sachs group in two space-time dimensions \cite{Afshar:2019axx}, describing the asymptotic symmetries of JT gravity in Bondi gauge. We have seen that the group is a semi-direct product, similarly to its higher-dimensional peers \cite{Bondi,Sachs2,Barnich:2014kra}. It can be defined with the same degree of rigour as the Virasoro group \cite{Guieu} or its warped generalization \cite{Detournay:2012pc,Afshar:2015wjm}. However, the central extensions involved in the definition and in the resulting algebra \eqref{bra1}--\eqref{bra3} are sharply different from those encountered so far in the literature, implying subtleties that do not normally affect discussions of JT gravity. In particular, the crucial presence of an extra zero-mode in the gravity action made it slightly different from the BMS-Schwarzian, in just the right way to reinstate the presence of saddle points along with translation symmetry. This eventually allowed us to evaluate the one-loop partition function of the theory.

One can think of this work as a first step towards the study of BMS$_2$ symmetry and its application to 2D gravity. Indeed, a number of issues have been omitted here and deserve deeper treatment. The complexification of BMS$_2$, for instance, is a seemingly abstract problem that happens to have crucial physical consequences: it explains the difference between partition functions \eqref{zfinal} and \eqref{zplane} from the presence of enhanced complex symmetries for (covers of) the Euclidean plane. It would be satisfactory to build a framework where such complex orbits appear naturally, as opposed to the rough arguments exposed in section \ref{serta}. Another aspect of BMS$_2$ symmetry that was not addressed here is its application to warped CFTs. Indeed, as  stressed repeatedly, BMS$_2$ extends the warped Virasoro group, so it is natural to wonder if it can be defined as a space-time symmetry of the kind that normally defines warped CFTs. One may then wonder if BMS$_2$ entails a Cardy-like formula similar to those of more standard extensions of the Virasoro group \cite{Bagchi:2012xr,Barnich:2012xq,Detournay:2012pc}. Applications to black hole entropy would then be conceivable, \eg along the lines of \cite{Afshar:2015wjm} or \cite{Afshar:2016uax,Afshar:2017okz}.

\section*{Acknowledgements}

We are grateful to Glenn Barnich and Charles Marteau for discussions on JT gravity and its asymptotic symmetries, and we especially thank Sergei Dubovsky, Daniel Grumiller and Romain Pascalie for insightful comments on an early version of this manuscript. The work of B.O.\ is supported by the ANR grant \textit{TopO} No.\ ANR-17-CE30-0013-01 and the European Union’s Horizon 2020 research and innovation programme under the Marie Sk{\l}odowska-Curie grant agreement No. 846244.

\appendix

\section{Adjoint representation of \texorpdfstring{BMS$_2$}{BMS2}}
\label{app1}

This appendix completes section \ref{sebalg} and describes the adjoint representation of the (centrally extended) BMS$_2$ group. This is a key technical step for the computation of Lie brackets \eqref{bak} and that of the BMS$_2$ coadjoint representation \eqref{ttj}--\eqref{ttq}.

The Lie algebra of the BMS$_2$ group is its tangent space at the identity, whose elements are triples $(\eps,\eta,\bz)\in\Vect\times\Omega^1(S^1)\times\RR^3$. It is acted upon by the adjoint representation, \ie the differential of conjugation at the identity:
\be
\Ad_{(f,\eta_1,\bz_1)}(\eps,\eta_2,\bz_2)
\equiv
\left.\frac{\di}{\di t}\right|_{t=0}
(f,\eta_1,\bz_1)\cdot
(e^{t\eps},t\eta_2,t\bz_2)\cdot
(f,\eta_1,\bz_1)^{-1}\,.
\ee
Owing to the group operation \eqref{muc}, the operator $\Ad_{(f,\eta,\bz)}$ is actually independent of central entries $\bz$, so we omit them in the subscript from now on. In fact, using \eqref{muc}, one finds
\be
\begin{split}
&\Ad_{(f,\eta_1)}(\eps,\eta_2,\bz)
=
\Big(%
f_*\eps,\,(f^{-1})^*\eta_2 +\cL_{f_*\eps}\eta_1;\,
\bz+\widetilde\bC[f,\eps,\eta_1,\eta_2]\Big)
\end{split}
\label{ad}
\ee
where $(f_*\eps)(\phii)\der_{\phii}\equiv\eps(f^{-1}(\phii))[(f^{-1})'(\phii)]^{-1}\der_{\phii}$ is the pushforward of the vector $\eps$ by the diffeomorphism $f$, while the three functionals in $\widetilde\bC[f,\eps,\eta_1,\eta_2]=\big(\widetilde\sfA[f,\eps],\widetilde\sfB[f,\eps,\eta_1,\eta_2],\widetilde\sfC[f,\eps,\eta_1,\eta_2]\big)$ respectively stem from the three cocycles \eqref{coc}--\eqref{coq} and are given by
\begin{align}
\widetilde\sfA[f,\eps]
&=
-\oint\frac{\di\phii}{24\pi}\,
\eps(\phii)\,\left(\frac{f'''}{f'}-\frac{3}{2}\left(\frac{f''}{f'}\right)^2\right)\,,\\
\label{wab}
\widetilde\sfB[f,\eps,\eta_1,\eta_2]
&=
\oint\frac{\di\phii}{2\pi}
\big[\eta_2\,\log f'+\eta_1'\,f_*\eps\big]\,,\\
\label{wac}
\widetilde\sfC[f,\eps,\eta_1,\eta_2]
&=
\oint\frac{\di\phii}{2\pi}
\big[\eta_2\,\big(f(\phii)-\phii\big)-\eta_1\,f_*\eps\big]\,.
\end{align}
It is understood here that all integrals run over the circle $\phii\in[0,2\pi)$, and we abuse notation by writing forms and vector fields and their components with the same symbols (for instance $(f_*\eps)(\phii)\equiv\eps(f^{-1}(\phii))[(f^{-1})'(\phii)]^{-1}$, etc.). Note that $\widetilde\sfA$ involves the usual Schwarzian derivative and also appears in the adjoint representation of the Virasoro group (see \eg \cite[eq.\ (6.98)]{Oblak:2016eij}), while the quantities \eqref{wab}--\eqref{wac} are new.

Eqs.\ \eqref{ad}--\eqref{wac} are implicitly used at two important points in the main text. First, the Lie bracket \eqref{bak} is obtained by differentiating the adjoint representation at the identity: schematically, $[\eps_1,\eps_2]\equiv\ad_{\eps_1}\eps_2\equiv-\left.\frac{\di}{\di t}\right|_{t=0}\Ad_{e^{t\eps_1}}\eps_2$ where the minus sign is included on the right-hand side for convenience. Generalizing this to triples $(\eps,\eta,\bz)$ and using \eqref{ad}--\eqref{wac} yields the result quoted in eq.\ \eqref{bak} above. Second, the coadjoint representation of eqs.\ \eqref{ttj}--\eqref{ttq} crucially relies on eqs.\ \eqref{ad}--\eqref{wac}, since it is the dual of the adjoint. Concretely, one schematically has $\langle\Ad^*_f\cT,\eps\rangle\equiv-\langle\cT,\Ad_{f^{-1}}\eps\rangle$, and generalizing to triples $(\eps,\eta,\bz)$ and $(\cT,\cQ,\bc)$ provides
\begin{align}
\label{toff}
\widetilde T
&=
[(f^{-1})']^2\,T\circ f^{-1}
-\frac{a}{12}\sfS[f^{-1}]
+b\,\eta\frac{(f^{-1})''}{(f^{-1})'}
-b\,\eta'
+c\,\eta(f^{-1})'
+\eta(Q\circ f^{-1})'\,,\\
\label{boff}
\widetilde Q
&=
Q\circ f^{-1}
+b\log[(f^{-1})']
+c\,(f^{-1}(\phii)-\phii)\,.
\end{align}
The first term on the right-hand side of \eqref{toff} is a CFT transformation law with central charge $a$ and involves the standard Schwarzian derivative, while the remaining terms are new. Evaluating both equations at $f(\phii)$ instead of $\phii$ yields eqs.\ \eqref{ttj}--\eqref{ttq} above.

\section{Symplectic form and measure}
\label{appE}

This appendix completes sections \ref{sehaar}--\ref{secomp} by displaying the construction of the Liouville/Haar measure on BMS$_2$. The technical core of the computation is the evaluation of the Maurer-Cartan form of the centrally extended BMS$_2$ group. Since the path integral of section \ref{secomp} localizes to the identity in BMS$_2$, we will focus on the value of the measure \textit{at} the saddle point. Furthermore, as explained in section \ref{secomp}, choosing a right-invariant measure allows us to limit ourselves to the simplest orbit representative $(\cT,\cQ)=(0,0)$ without loss of generality.

The strategy is as follows. We first evaluate the Maurer-Cartan form of BMS$_2$, then pair it with a coadjoint vector $(0,0;a,b,c)$, and finally take the exterior derivative in field space to read off the Kirillov-Kostant symplectic form. The corresponding path integral measure then is the infinite power of the symplectic form and involves the Pfaffian described in section \ref{secomp}.

The right Maurer-Cartan form of BMS$_2$ follows from the centrally extended group operation \eqref{muc}, and thus generalizes the centreless Maurer-Cartan form \eqref{e44}. To evaluate it, let $(f_t,\eta_t)$ be some path in BMS$_2$, defining a tangent vector $(\dot f_0,\dot\eta_0)$ at $t=0$. (For simplicity we exclude central components from the path; these components do not contribute to the measure anyway.) One then finds from \eqref{muc} that the right logarithmic derivative of the path is
\be
\label{B1}
\begin{split}
\der_{\tau}\Big|_0
\left[(f_{\tau},\eta_{\tau};0)\cdot(f_t,\eta_t;0)^{-1}\right]
=&
\Bigg(%
\dot f\circ f^{-1},
\dot\eta+(\eta\,\dot f\circ f^{-1})',
\oint\frac{\di\phii}{48\pi}(\log f')'\partial_t(\log f'),\\
&
\quad
-\oint\frac{\di\phii}{2\pi}\dot f'\,\eta\circ f,
-\oint\frac{\di\phii}{2\pi}\dot f\,f'\,\eta\circ f
\Bigg)\,.
\end{split}
\ee
The first two entries on the right-hand side reproduce the centreless result \eqref{e44}, while the central entries are new and involve among others the standard Virasoro geometric action $\propto\oint\phi'\dot\phi$ in terms of $\phi=\log f'$ \cite{Alekseev:1988ce}. The corresponding Lie algebra-valued one-form $\Theta$ is, by definition, the right Maurer-Cartan form on the (centrally extended) BMS$_2$ group, obtained by trading all time derivatives in \eqref{B1} for functional exterior derivatives (as already done in \eqref{e44}):
\be
\label{bethet}
\Theta
=
\Bigg(%
\delta f\circ f^{-1},
\delta\eta+(\eta\,\delta f\circ f^{-1})',
\oint\tfrac{\di\phii}{48\pi}(\log f')'\delta(\log f'),
-\oint\tfrac{\di\phii}{2\pi}\delta f'\,\eta\circ f,
-\oint\tfrac{\di\phii}{2\pi}\delta f\,f'\,\eta\circ f
\Bigg)\,.
\ee
The Kirillov-Kostant symplectic form $\Omega$ evaluated at the point $(\cT,\cQ;a,b,c)$ is then obtained by pairing the latter with the Maurer-Cartan form and taking the exterior derivative: $\Omega\equiv\delta\langle(\cT,\cQ;a,b,c),\Theta\rangle$ in terms of the adjoint-coadjoint pairing \eqref{copa}. (See \eg \cite[sec.\ 5.3.2]{Oblak:2016eij} for the details of this procedure.) As announced above, we take $(\cT,\cQ)=(0,0)$ without loss of generality so that only the central entries of \eqref{bethet} contribute, which yields
\be
\Omega
=
-\frac{a}{24}\oint\frac{\di\phii}{2\pi}\frac{\delta f'\wedge\delta f''}{f'^2}
-b\oint\frac{\di\phii}{2\pi}\delta(\eta\circ f)\wedge\delta f'
+c\oint\frac{\di\phii}{2\pi}\delta f \wedge\delta\,(f'\eta\circ f)\,.
\label{symplecticformBis}
\ee
The first term is the well-known central part of the symplectic structure of Virasoro coadjoint orbits \cite{Witten:1987ty,Alekseev:1988ce}, while the second and third terms are specific to BMS$_2$. This is the expression used in section \ref{secomp} to evaluate the Pfaffian \eqref{pfaffian} that is eventually crucial for the value of the partition function.

Note that \eqref{symplecticformBis} reduces to the symplectic form of warped Virasoro orbits \cite{Afshar:2019tvp} when $\eta=\sigma'$ is exact. Indeed, the second and third terms on the right-hand side of \eqref{symplecticformBis} then become
\be
-\oint\frac{\di\phii}{2\pi}\frac{\delta\sigma'}{f'}\wedge\delta f'
     +\cpz\oint\frac{\di\phii}{2\pi}\delta f\wedge\delta\sigma'\,,
\ee
which coincides indeed with the symplectic form of warped Virasoro coadjoint orbits at vanishing U(1) level \cite{Afshar:2019tvp}. This is consistent with the embedding of warped Virasoro in BMS$_2$ emphasized throughout this work.

\addcontentsline{toc}{section}{References}

\providecommand{\href}[2]{#2}

\end{document}